\documentclass[11pt,preprint]{aastex}
\newcommand{\etal}{~{et~al.}\ } 
\newcommand{\unit}[1]{\ifmmode\,{\rm #1}\else$\,{\rm #1}$\fi}
\newcommand{\mujy}{\unit{\mu Jy}}
\newcommand{\phno}{$\phantom{0}$}

\usepackage{ulem}
\usepackage{natbib}
\slugcomment{Accepted by AJ, July 11, 2005}

\shorttitle{Lensed Radio Lobe Survey}
\shortauthors{Haarsma et al.}

\begin{document}

\title{The FIRST-Optical-VLA Survey for Lensed Radio Lobes}

\author{
Deborah B.\ Haarsma\altaffilmark{1},
Joshua N.\ Winn\altaffilmark{2,3},
Emilio E.\ Falco\altaffilmark{4},
Christopher S.\ Kochanek\altaffilmark{5},
Philip Ammar\altaffilmark{1},
Catherine Boersma\altaffilmark{1},
Shannon Fogwell\altaffilmark{1},
T. W. B. Muxlow\altaffilmark{6},
Brian A.\ McLeod\altaffilmark{2},
Joseph Leh{\'a}r\altaffilmark{2,7}
}

\altaffiltext{1}{Calvin College, 1734 Knollcrest SE, Grand Rapids, MI 49546, dhaarsma@calvin.edu}

\altaffiltext{2}{Harvard-Smithsonian Center for Astrophysics, 60
Garden Street, Cambridge, MA 02138}

\altaffiltext{3}{Hubble Fellow}

\altaffiltext{4}{F.L.\ Whipple Observatory, Smithsonian Institution,
P.O.\ Box 97, Amado, AZ 85645}

\altaffiltext{5}{Department of Astronomy, The Ohio State University,
4055 McPherson Lab, 140 West 18th Avenue, Columbus, OH 43210}

\altaffiltext{6}{University of Manchester, Jodrell Bank Observatory,
Macclesfield, Cheshire SK11 9DL, UK}

\altaffiltext{7}{Current address: CombinatoRx, Inc., 650 Albany St.,
Boston, MA 02118, and Boston University Bioinformatics, 44 Cummington
St., Boston, MA 02115}

\begin{abstract}
We present results from a survey for gravitationally lensed radio
lobes. Lensed lobes are a potentially richer source of information
about galaxy mass distributions than lensed point sources, which have
been the exclusive focus of other recent surveys. Our approach is to
identify radio lobes in the FIRST catalog and then search optical
catalogs for coincident foreground galaxies, which are candidate
lensing galaxies. We then obtain higher-resolution images of these
targets at both optical and radio wavelengths, and obtain optical
spectra for the most promising candidates. We present maps of several
radio lobes that are nearly coincident with galaxies. We have not
found any new and unambiguous cases of gravitational lensing. One
radio lobe in particular, FOV~J0743+1553, has two hot spots that could
be multiple images produced by a $z=0.19$ spiral galaxy, but the
lensing interpretation is problematic.
\end{abstract}


%%%%%%%%%%%%%%%%%%%%%%%%%%%%%%%%%%%%%%%%%%%%%%%%%%%%%%%%%%%%%%%%%%%%%%%%%%
\section{Introduction}

Gravitational lenses are prized for their ability to tell us about
galaxies.  The configuration of the multiple images of an
extragalactic background source provides a direct and accurate
measurement of the enclosed mass of the foreground galaxy, including
the contribution due to dark matter (as reviewed by Narayan \&
Bartelmann 1999; Courbin, Saha, \& Schechter 2002; Kochanek\etal
2004).\nocite{narayan99a,courbin02a,kochanek04b} Lenses can be used to
study the mass distributions of individual galaxies
(\citealp{kochanek91a}; for recent examples see Trotter, Winn, \&
Hewitt 2000\nocite{trotter00a}; \citealp{cohn01a,rusin02a}), as well
as statistical properties of galaxy mass distributions such as the
degree of alignment of mass and light (e.g., Keeton, Kochanek, \&
Seljak 1997\nocite{keeton97b}), the radial mass distribution (e.g.,
Davis, Huterer, \& Krauss 2003\nocite{davis03a}; Rusin, Kochanek, \&
Keeton 2003b\nocite{rusin03b}), and the degree of central
concentration \citep{rusin01a}.  One the most successful applications
of lenses has been to measure the evolution of the mass-to-light ratio
of field galaxies with cosmic epoch (\citealp{kochanek00a, chae03b,
rusin03a}; van de Ven, van Dokkum, \& Franx 2003\nocite{vandeven03a};
Ofek, Rix, \& Maoz 2003\nocite{ofek03a}).

Our ability to use lenses in this manner is limited by the relatively
small number of known lenses and the small number of constraints that
each one imposes on the mass distribution (see \citealp{kochanek04b}).
We have undertaken a survey (which we term the
   FIRST\footnote{Faint Images of the Radio Sky at Twenty-Centimeters,
   http://sundog.stsci.edu}-Optical-VLA\footnote{The Very Large Array
   (VLA) is operated by the National Radio Astronomy Observatory
   (NRAO), a facility of the National Science Foundation operated
   under cooperative agreement by Associated Universities for Research
   in Astronomy (AURA), Inc.}, or FOV, survey),
the goal of which is not only to discover more lenses, but to discover
lenses in which the background source is extended rather than
pointlike. For the purpose of studying the mass distribution of the
foreground galaxy, gravitationally lensed extended structures provide
a potentially richer set of constraints than lensed point sources.

Many of the approximately 90 known cases of multiple-image lensing by
galaxies were discovered in systematic lens surveys. Of these, roughly
equal numbers were discovered by imaging surveys of radio sources
[e.g.\ MG--VLA \citep{burke93a}, JVAS/CLASS \citep{browne03a}, and
PMN--NVSS \citep{winn01b,winn02a}] and optically selected quasars
  [HST
% in acknowledgements:
%\footnote{The NASA/ESA Hubble Space Telescope (HST) is operated
%  by AURA, Inc. under NASA contract NAS 5-26555.}
%
snapshot surveys \citep{maoz93a,gregg00a} and
ground based surveys such as WFI \citep{morgan04a}, the Hamburg--ESO
survey \citep{wisotzki04a}, and the 
  SDSS\footnote{Funding for the Sloan Digital Sky Survey (SDSS)
  has been provided by the Alfred P. Sloan Foundation, the
  Participating Institutions, the National Aeronautics and Space
  Administration, the National Science Foundation, the U.S. Department
  of Energy, the Japanese Monbukagakusho, and the Max Planck
  Society. The SDSS Web site is http://www.sdss.org .}
\citep{oguri05a}].  Optical surveys have generally resulted in a larger
number of lenses per source examined, because of a greater
magnification bias (the tendency for lensed quasars to be
over-represented in a flux-limited sample) than radio-selected
samples.  On the other hand, radio surveys have the advantages of the
superior angular resolution and high degree of automation of the VLA,
and freedom from extinction by dust.  Radio surveys also differ from
optical surveys in that the background sources are not necessarily
point sources, as are optically-selected quasars. The classical
``double-lobed radio galaxy'' has a compact core of radio emission
near the center of the galaxy, and thin jets of radio emission leading
to more diffuse lobes that can be located hundreds of kiloparsecs
away.  The relative intensities of the core, jet, and lobe components
varies considerably from one source to another, and also with
observing frequency.

The largest and most recent radio surveys (JVAS/CLASS and PMN--NVSS)
concentrated on finding lensed radio cores, rather than lobes.  The
cores are typically point-like at arcsecond resolution, and can be
selected from catalogs with poorer resolution by virtue of their flat
radio spectra, whereas lobes are typically diffuse and have
  steep radio spectra\footnote{Here, ``flat'' means that the radio
  flux density is a weak function of observing frequency, and
  ``steep'' means that the flux density declines sharply with
  frequency. Supposing $S_\nu \propto \nu^{-\alpha}$, the cutoff
  between flat and steep is often taken to be $\alpha\approx 0.5.$}.
These surveys have been very successful, finding about one lens per
700 targets imaged by the VLA.  There are several reasons why
flat-spectrum sources were preferred.  First, the simple structure of
the cores makes lenses relatively easy to recognize: they have
multiple compact components separated by $\sim$1\arcsec. Second, the
high surface brightness of cores facilitates the interferometric
imaging process (i.e., self calibration and deconvolution). Third, the
radio cores or their optical counterparts sometimes have time-variable
fluxes, allowing the determination of time delays and thus a
combination of the Hubble constant and the surface density of the lens
(see, e.g., \citealp{kochanek03a}).  Fourth, the measured lensing
probability, which can be used to estimate the cosmological model
\citep{turner90a,fukugita92a}, is easier to interpret in a survey of
point sources than a survey of extended structures.

The main disadvantage of flat spectrum surveys, as already mentioned,
is that lensed point sources do not supply many constraints on the
mass distribution of the lens galaxy.  The simple two-image and
four-image lenses that are the primary yield of the flat spectrum
surveys provide estimates of the mass and the mean quadrupole of the
lens galaxy, but little else.  Furthermore, the flat-spectrum surveys
have already examined most of the sources in the sky that are
sufficiently bright to be observable with the very short VLA
observations (about 30~seconds) that are necessary to make up for the
low lensing probability. This situation will change with the
development of the
  EVLA\footnote{Expanded Very Large Array (ELVA) http://www.aoc.nrao.edu/evla} 
and
  e-MERLIN\footnote{e-MERLIN http://www.merlin.ac.uk/e-merlin}, 
but these capabilities are still some years away.

Our survey pursues the population of extended radio sources: jets and
lobes that are lensed into partial or complete Einstein rings. In
fact, these steep-spectrum lenses outnumber the flat-spectrum lenses
on the sky. Flat and steep spectrum sources are similar in number
density and redshift distribution, but the greater angular extent of
the steep spectrum sources greatly enhances the probability that a
foreground galaxy is close enough to the line of sight for strong
lensing to occur \citep{kochanek90a}. We expect roughly 75\% of radio
lenses to be steep spectrum sources. Indeed, many of the lenses found
by the MG--VLA image survey, which targeted both flat and steep
spectrum radio sources, were Einstein rings: MG~1131+0456,
\citep{hewitt88b}, MG~1654+1346 \citep{langston89a}, MG~1549+3047
\citep{lehar93a}, and MG~0751+2716 \citep{lehar97a}.  The Einstein
rings PKS~1830-211 \citep{jauncey91a} and B~0218+357
\citep{patnaik93a} were found in flat spectrum surveys, although the
rings in these systems are much fainter than the cores.  The challenge
of steep-spectrum surveys, as discovered by the MG--VLA survey, is
that the complex structure of the radio sources means that more
follow-up observations are required to separate the lenses from
unlensed lobes that happen to have intrinsically ring-like morphology.
Nonetheless, it is worth finding the steep spectrum lenses---many
applications of lenses are still limited by the available numbers of
systems, and the extended structure of the steep spectrum lenses
provide far more constraints on mass distributions.

We have developed a more efficient method to find steep-spectrum
lenses \citep{lehar01a}.  Rather than starting with a complete catalog
of radio sources (as the MG--VLA survey), we pre-select sources that
are likely to be radio lobes, by employing a source catalog with
sufficiently good angular resolution.  Then we search for optical
objects at the same position as the lobes.  Because radio lobes
themselves have no optical emission, any galaxy that is detected is
likely to be a foreground galaxy, and therefore a potential lens
galaxy.  In particular, we combine the relatively high resolution
FIRST survey with relatively deep optical surveys
  (APM\footnote{Cambridge Automated Plate Measurement (APM) of POSS-1
  and UKST plates http://www.ast.cam.ac.uk/$\sim$apmcat}
and SDSS). This strategy proved successful in our pilot study
\citep{lehar01a} where we used this method on a portion of the FIRST
catalog, finding 2 lenses (the new lens FOV~J0816+5003, and a
re-discovery of MG~1549+3047) among a list of only 33 candidates.

Here we describe an extension of the FOV survey to a larger portion of
the FIRST catalog. In \S\ref{target.selection} we describe the
selection of targets from radio and optical catalogs.  In
\S\ref{radio.obs} and \S\ref{opt.obs} we present follow-up radio and
optical observations of these targets, which greatly narrowed the
list of lens candidates. In \S\ref{J0743} we describe the particular
candidate FOV~J0743+1553 in some detail, since the initial
observations appeared to be promising.  We discuss the other
candidates in \S\ref{other}, and summarize our conclusions in the
final section. In this paper, we assume a flat, $\Omega_m=0.3$
cosmology with $h=0.72$ whenever cosmological calculations are needed.

%%%%%%%%%%%%%%%%%%%%%%%%%%%%%%%%%%%%%%%%%%%%%%%%%%%%%%%%%%%%%%%%%%%%%%%
\section{Target Selection}
\label{target.selection}

The target selection process had four steps: selection of a
flux-limited sample of radio sources; identification of likely radio
lobes; identification of any optical counterparts to the lobes; and
final visual inspection to filter out catalog errors and prioritize
candidates for further follow-up. In this section we describe each of
these steps in more detail.

The flux-limited sample was drawn from the VLA FIRST catalog of radio
sources (Becker, White, \& Helfand 1995)\nocite{becker95a}.  This
20~cm (1.4~GHz) survey is complete to $\sim$1~mJy, with a beam size of
5\farcs4 FWHM.  While the angular resolution is not fine enough to
detect the arcsecond-scale separations between lensed images, it does
suffice to identify objects that are likely to be radio lobes.  For
our survey we consulted the 2001~October~15 edition of the catalog,
which contained 771,076 radio objects in 7954~square degrees in the
north Galactic cap and 611~square degrees in the south Galactic cap.
We selected objects brighter than 3~mJy that were not flagged as radio
sidelobes (375,919 objects).

Next, we identified likely radio lobes.  Following the algorithm
presented in the appendix of \cite{lehar01a}, we identified groups of
two or more FIRST sources that were spaced closely enough to represent
the core and lobe (or lobes) of a single radio galaxy. In particular
we selected groups in which the maximum separation between any two
sources is less than 90\arcsec, and in which no two sources are
separated by more than 60\arcsec (these numbers were chosen with
reference to the typical size of radio galaxies). For groups with
three or more sources, the algorithm selected those that are more
symmetric (symmetry factor $S>0.50$) and less bent (bending angle
$\theta<30$\degr) (see the appendix of \citealp{lehar01a} for definitions
of $S$ and $\theta$).  The algorithm then identified which components
in the group are likely to be lobes and which component is likely to
be the core (using the extent of the objects and their location within
the group).  Of these groups, we selected those which would be most
readily observable: we required that at least one object in the group
be bright enough for self-calibration ($>$10~mJy at 20~cm), and that
the lobes be small enough to provide sufficient contrast in
high-resolution follow-up observations ($<$10\arcsec). This gave us
30,927 radio galaxies with identifiable, bright lobes.

To find optical objects at the same position as the radio lobes, we
used the Cambridge APM scans of the POSS-I and UKST plates. The POSS-I
plates cover the northern sky to a limiting magnitude of 21.5 in O
(blue) and 20 in E (red).  The southern UKST plates have a limiting
magnitude of 22.5 in B and 21 in R.  The APM objects near FIRST
objects were identified for the entire 2001~October~15 FIRST catalog
(see \citealp{mcmahon02a}).  We searched for all APM objects within
5\arcsec\ of the radio lobes on our list, and also noted any
counterparts to the radio cores.  In addition, we searched the SDSS
Early Data Release \citep{stoughton02a}, which covered only 462 square
degrees but was complete to 22.2 in $r$, in a similar manner.  These
searches identified 4500 lobes with coincident optical objects.  Since
this number was too large for a realistic follow-up campaign, we
further narrowed the list to 380 candidates by requiring the radio
lobes be brighter than 100~mJy at 20~cm, which allowed for the most
efficient follow-up radio observations.

To make the final cut, three of us (D.B.H., J.L., J.N.W.) visually
inspected the targets selected by the computer algorithm.  We checked
that the FIRST radio morphology was consistent with a double-lobed
radio galaxy, and that the optical object appeared to be real (rather
than an image artifact) and was indeed well aligned with a lobe.
Objects were deemed better targets if the optical counterpart was red
or extended (making it more likely to be a galaxy, rather than a
foreground star), and if a separate optical object was coincident with
the radio core (since this would facilitate redshift determination of
the radio galaxy).  With these criteria we selected 92 targets for
high-resolution follow-up VLA observations.

%%%%%%%%%%%%%%%%%%%%%%%%%%%%%%%%%%%%%%%%%%%%%%%%%%%%%%%%%%%%%%%%%%%%%%%
\section{Radio Observations}
\label{radio.obs}

We obtained high-resolution VLA images for all 92 targets. The goals
of this step of the survey were to confirm that the FIRST source
really is a radio lobe, to obtain a more accurate radio position for
comparison with the catalog position of the optical object, and to
look for lensed structures (e.g., rings, arcs, and multiple images) on
the expected scale of $\sim$1\arcsec.  Table~\ref{table.vla} gives the
coordinates, exposure time, and observing wavelength for each of the
VLA observations.  Of these, 11 had been previously observed and were
available in the VLA archive or the literature. The remaining 81
targets were observed over 32.5 hours in four sessions in March
2002. These observations took place when the VLA was in the A
configuration, and employed the standard 6~cm (5~GHz) band, giving a
synthesized beam size of $\sim$0\farcs4 FWHM.  A bandwidth of 100~MHz
was recorded for each circular polarization.  The data were calibrated
with standard
  AIPS\footnote{The Astronomical Image Processing System (AIPS) is
  developed and distributed by the National Radio Astronomy
  Observatory.}.
procedures. The flux density scale was set using 3C~286 and 2355+498,
and the polarization was calibrated using 1310+323 and 2355+498. The
data were mapped and deconvolved using the standard CLEAN algorithm,
and two rounds of self-calibration were applied to the antenna phases.

In the resulting maps, 9 targets were unresolved, suggesting that they
are actually radio cores rather than lobes.  Another 53 targets were
indeed lobes, but the more accurate radio position showed that the
lobe emission did not coincide with the catalog position of the
optical object.  The remaining 30 candidates were re-mapped using the
multi-scale CLEAN (MSC) deconvolution algorithm \citep{wakker88a}
and are shown in Figure 1.  The standard CLEAN algorithm
tends to introduce point-like artifacts (a ``pebbly'' look) into
extended radio sources, because it models the sky as a set of point
sources, which is a poor approximation for diffuse radio lobes.  The
MSC algorithm assumes the sky is composed of both points and
elliptical Gaussian components of various dimensions.  Using MSC, we
were able to reduce the artifacts in the maps and provide a more
realistic description of the actual radio source. The rms noise in the
MSC-deconvolved maps ranged from 35 to 90\mujy~beam$^{-1}$, and
was usually less than 60\mujy~beam$^{-1}$.

In at least 24 of these 30 cases, there was a radio point source
positioned between the lobes, presumably the core of the radio galaxy.
An optical counterpart to this radio point source was present in 18
cases, which was used for more precise registration of the radio and
optical images.

%%%%%%%%%%%%%%%%%%%%%%%%%%%%%%%%%%%%%%%%%%%%%%%%%%%%%%%%%%%%%%%%%%%%%%%
\section{Optical Observations}
\label{opt.obs}

In the next phase of the survey, we obtained CCD optical images of all
30 candidates, in order to confirm the existence of the optical
counterpart and eliminate spurious catalog entries, to obtain a more
accurate optical position and test whether it truly coincides with the
radio lobe, and to confirm that it is a galaxy rather than a star by
measuring its angular extent.  Table~\ref{table.opt} summarizes the
follow-up optical observations, which took place from 2002 through
2004, using a variety of telescopes and cameras.  In some cases we
obtained images of a single candidate using more than one telescope,
but in what follows we describe only the images with the best seeing
and signal-to-noise ratio.

Fourteen candidates were imaged with the 4-shooter CCD camera on the 
  FLWO 1.2~meter telescope\footnote{The Fred Lawrence Whipple
  Observatory (FLWO) is a facility of the Smithsonian Institution.},
for 1200 seconds in each of V and I bands.  Two candidates were imaged
with the MagIC camera on the
  6.5~m Baade Magellan Telescope\footnote{The 6.5~m Baade telescope is
  one of the two telescopes of the Magellan Project, a collaboration
  between Carnegie Observatories, the University of Arizona, Harvard
  University, the University of Michigan, and the Massachusetts
  Institute of Technology.}
in the g$'$ and i$'$ bands.  Seven candidates were imaged 
at the 
  MDM 2.4~meter Hiltner Telescope\footnote{The MDM Observatory is
  owned and operated by University of Michigan, Dartmouth College,
  Ohio State University, and Columbia University.}
in the I band for 900 to 1800 seconds each.  Finally, for 7
candidates, the best available CCD images were drawn from the SDSS
Data Release 1 (DR1) and Data Release 2 (DR2), which were made public
after our VLA observations. Among the SDSS data products is a
classification of each object as either pointlike or extended, which
is deemed reliable for objects with r magnitude brighter than 21. We
provide these results, where available, in Table~\ref{table.opt} and
Figure 1.

In those cases for which the core was detected at both optical and
radio wavelengths, the images were registered by requiring coincidence
of the optical and radio core.  If this was not possible, then the VLA
radio position was matched to the SDSS optical position (or a plate
solution using
  USNO-B1.0 stars\footnote{The USNOFS Image and Catalogue Archive is
  operated by the United States Naval Observatory, Flagstaff Station,
  http://www.nofs.navy.mil/data/fchpix/.},
for cases in which the SDSS position was not available).

Because two of the 30 candidate radio galaxies have an optical
counterpart to both lobes, there are a total of 32 candidate lens
galaxies in our sample.  Nine of the 32 candidates were eliminated
through optical follow-up: one is a spurious catalog entry, three are
not coincident with the radio lobe, and five are stellar.  Most of the
remaining lens candidates have non-stellar optical counterparts that
are well aligned with a radio lobe (see Table~\ref{table.status}).

In a few cases, redshifts of either the radio galaxy or the candidate
lens galaxy were already known.  We searched the 
  NASA/IPAC Extragalactic Database\footnote{The NASA/IPAC
  Extragalactic Database (NED) is operated by the Jet Propulsion
  Laboratory, California Institute of Technology, under contract with
  the National Aeronautics and Space Administration.},
finding redshifts for 5 core counterparts and 3 candidate lens
galaxies.  These redshifts are given in Table~\ref{table.status} and
Figure 1.  We also searched SDSS DR1 and DR2 for redshifts of our
candidate objects, but we did not find any additional information.

%%%%%%%%%%%%%%%%%%%%%%%%%%%%%%%%%%%%%%%%%%%%%%%%%%%%%%%%%%%%%%%%%%%%%%%
\section{Target FOV~J0743+1553}
\label{J0743}

\begin{figure}
\figurenum{1a}
\plotone{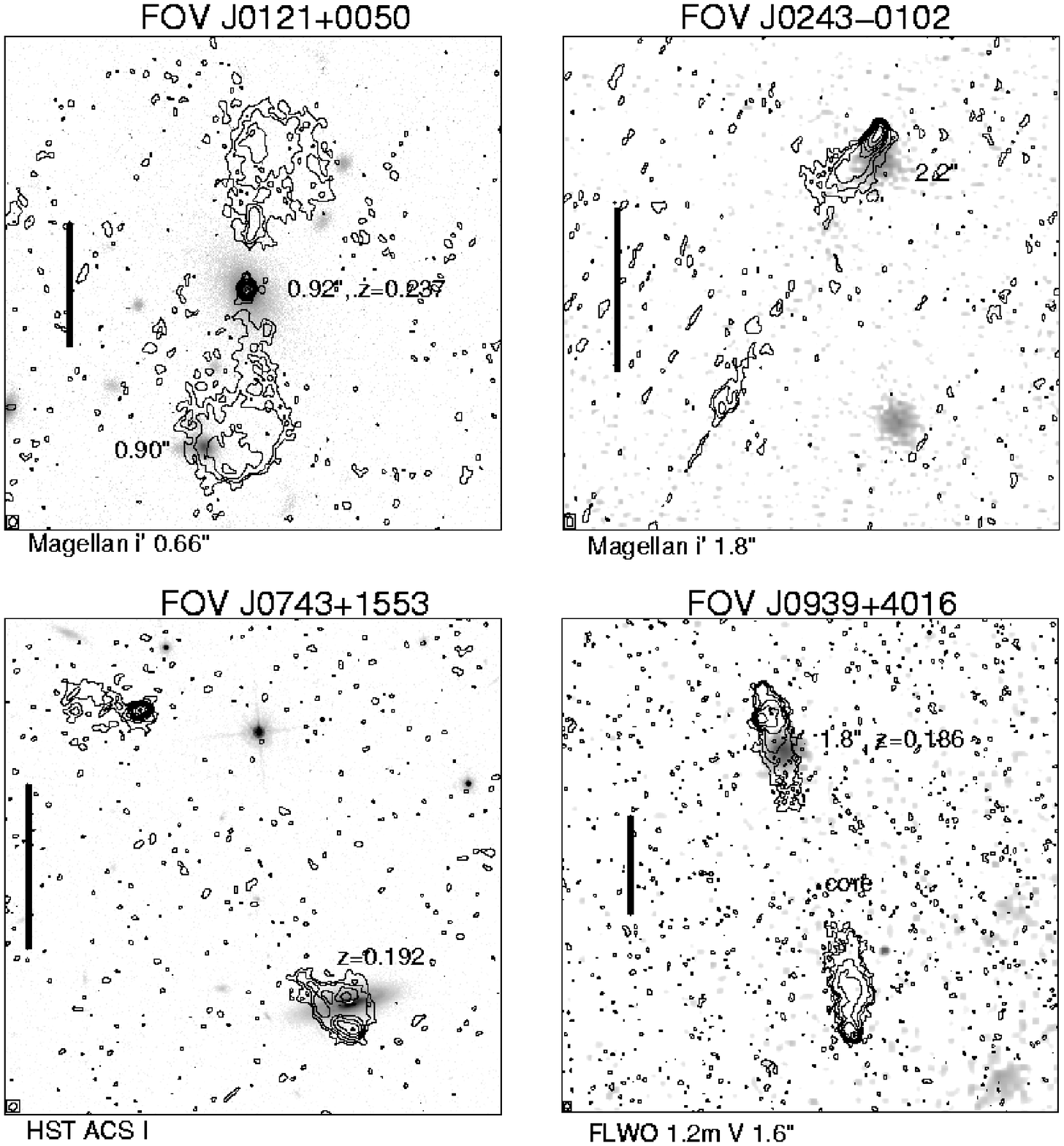} %lenses1.eps
\caption{Radio and optical maps of the top 30 candidates.  North is up
and east is to the left.  The vertical bars are 10 arcseconds long.
The contours are VLA data, with contour levels increasing by doubling
from twice the off-source map rms; the beam is shown in the lower
left.  The optical data is shown in a square-root grey scale selected
to highlight the potential lens and/or core counterpart, and the
telescope, filter, and seeing is given in the lower left. The measured
FWHM extent of certain optical objects is indicated in the images.
For SDSS data, designations of ``point'' and ``extended'' are shown
for brighter objects ($r<21$).  Crosses indicate the position of
optical catalog objects which are not clearly visible in the optical
image.
\label{fig.lenses1} }
\end{figure}

The results for all 32 candidate lens galaxies are discussed in the
next section.  One of the targets, FOV~J0743+1553, captured our
interest because at 6~cm the morphology is suggestive of a
doubly-imaged hot spot (Figure~\ref{fig.lenses1}). Furthermore, the
optical counterpart proved to be a spiral galaxy, raising the prospect
of the discovery of a rare case in which a spiral galaxy acts as a
gravitational lens.  Our follow-up observations are described below,
along with a description of the strengths and weaknesses of the
lensing hypothesis.

%----------------------
\subsection{A spiral galaxy} 

Ground-based images revealed the optical counterpart to FOV~J0743+1553
to be highly elliptical and suggestive of a disk galaxy. We obtained
images with higher angular resolution using the {\it Hubble
Space Telescope} ({\it HST}), under the auspices of the
  CASTLES\footnote{The CfA--Arizona Space Telescope Lens Survey
  (CASTLES) \citealp{castles}} 
project.  Optical images (using filters F814W and F555W) were obtained
on 2003~September~10 with the Advanced Camera for Surveys (ACS).
Near-infrared images were obtained on 2003~October~6 with the Near
Infrared Camera and Multi-Object Spectrometer (NICMOS) using the F160W
filter.  The 4 dithered exposures obtained through each filter were
combined, using the MultiDrizzle software \citep{koekemoer02a} for the
ACS exposures and the NICRED software \citep{lehar00b} for NICMOS
exposures.  The final images are displayed in
Figures~\ref{fig.J0743HSTH} and \ref{fig.J0743HSTV+I}.

These images confirmed that the optical counterpart is a spiral
galaxy. The galaxy is highly elongated and has pronounced dust lanes
that are most obvious in the bluest (F555W) image. The galaxy is
viewed somewhat edge-on; given its aspect ratio, the inclination is
approximately 60 degrees. We found that the H band (F160W) surface
brightness distribution is well described by a combination of two
analytic functions: a de~Vaucouleurs profile, representing the bulge,
and an exponential profile, representing the disk.  The parameters of
the best-fitting model of the surface brightness are given in
Table~\ref{table.optmodel}.

We also obtained an optical spectrum of the galaxy using the 6.5m
telescope at the
  MMT Observatory\footnote{The MMT Observatory is a joint facility of
  the Smithsonian Institution and the University of Arizona.}
in 2002~April.  The spectrum, reduced in the standard manner with
IRAF, is shown in Figure~\ref{fig.J0743spect}.  There are emission
lines of \ion{o}{3}, H$\alpha$ and \ion{o}{2}, as well as absorption
features due to hydrogen and alkali metals, all at a redshift of
$0.1918\pm0.0002$.  The nebular lines are consistent with a spiral
galaxy.

The rarity of spiral lenses makes the prospect of discovering a new
spiral lens especially appealing. The sample of known lensing galaxies
is overwhelmingly dominated by massive elliptical galaxies.  Spiral
galaxies have a smaller lensing cross-section than elliptical galaxies
and are less likely to be found as lenses. Of the roughly 90 lens
systems listed by CASTLES, only five are known to be spiral galaxies
(see Winn, Hall, \& Schechter 2003\nocite{winn03c} for a summary of
spiral lenses).  For several of the known spiral galaxy lens systems,
accurate photometry of the lens galaxy is difficult because the field
is very crowded or the quasar images are very bright, but
FOV~J0743+1553 does not suffer from those problems.  Unfortunately,
the hypothesis that spiral galaxy FOV~J0743+1553 is a case of lensing
is problematic.  In what follows, we summarize the evidence for and
against this hypothesis.

%--------------------------------
\subsection{The case for lensing}

The morphology of the radio source at 6~cm is compatible with the
lensing interpretation. The southern lobe, shown in
Figure~\ref{fig.J0743VLA}), consists of a faint ring and several
brighter components. Two components of the ring, labeled A and B, are
1\farcs8 apart in a roughly north--south direction. These can be
interpreted as two lensed images of the same region of radio emission
in the background lobe. These components bracket the position of the
foreground galaxy. The fainter component is closer to the galaxy
centroid, as expected from lensing theory.  Furthermore, components A
and B have similar fractional polarizations, and similar polarization
angles. These properties are natural consequences of gravitational
lensing, but are not necessarily expected of two distinct hot spots in
a radio lobe. The other parts of the southern lobe---the bright region
on the southwest side of the lobe, and the jet extending to the core
to the northeast---have different polarizations, and are thus unlikely
to represent multiple images of a single source.

It is possible to make a simple model for the mass distribution of the
lens galaxy and the intrinsic radio source structure that is
compatible with the 6~cm data. The dark matter halo of the spiral
galaxy is modeled as a singular isothermal sphere with an Einstein
radius of $\sim$1\farcs8. The model is shown in
Figure~\ref{fig.J0743model}. The upper right panel shows the model of
the intrinsic 6~cm radio source structure (a sum of 7 elliptical
Gaussian functions), while the upper left panel shows the lensed
source. The center of the mass distribution was adjusted (within the
error uncertainty of the radio--optical registration) until the model
provided a good visual fit to the observed radio morphology; the mass
is centered 0\farcs03 east and 0\farcs45 south of component B. This
model is offered as an ``existence proof,'' rather than a unique
solution.  It demonstrates that a foreground mass would be able to
lens a portion of the lobe into images A and B without producing
multiple images in other parts of the lobe.  The hypothesized source
structure is reasonable; it is reminiscent of many radio jets and
lobes.

%--------------------------------
\subsection{The case against lensing}

Radio lobes show a great diversity in morphology. For this reason, a
lens-like morphology is not sufficient as proof that gravitational
lensing is occurring.  In contrast, the situation with flat-spectrum
sources is often clearer, especially in the case of quadruply-imaged
quasars, since there are no other known types of radio source that
exhibit 4 compact flat-spectrum components. In radio lobes, hot spots
and filaments can be arranged to look like multiple images or rings,
even when no lens galaxy is detected in optical or infrared. Two
examples in the literature are 4C~39.24 \citep{law-green95a} and
MG~0248+0641 \citep{conner98a}; some additional examples are given in
\S~\ref{other} (FOV~J1508+0102 and FOV~J1613+3724).  The bottom line
is that additional evidence beyond morphology is particularly
important.

The lensing hypothesis requires that the radio source be behind the
spiral galaxy.  Unfortunately, the redshift of this radio galaxy
cannot be measured, because the optical counterpart to the compact
radio core has not been detected (the radio core is northeast of the
lobe, Figure~\ref{fig.lenses1}).  Our most constraining upper limit on
the brightness of an optical point source is $I>26$, from the {\it
HST}\, ACS image.  Statistically, FIRST radio lobes tend to have
redshift $z\sim 1$ \citep{cress98a}, which is greater than the
redshift of the spiral galaxy ($z=0.19$).  But without a redshift
measurement of this particular radio source, we cannot confirm whether
it is in the background or the foreground of the spiral galaxy.  This
certainly reduces the confidence one can place in any particular
interpretation of the system.  The lack of a detectable optical
counterpart also inhibits the accuracy of the optical--radio
registration; we have registered the optical and radio images using
astrometry from the VLA observation and stars from the USNO-B1.0
catalog.

Lensing preserves the spectrum of a background object. For this
reason, the multiple images of a gravitational lens will have the same
arrangement and relative flux densities regardless of the observing
  frequency\footnote{We note that in some relatively rare cases,
  differential propagation effects such as scatter-broadening and
  free-free absorption can complicate this test of the lensing
  hypothesis (see, e.g., \citet{sykes98a,biggs04a,winn04a}).},
whereas the intrinsic source structure of radio lobes can vary
dramatically with frequency.  We observed FOV~J0743+1553 at 18~cm in
order to test whether the lens-like morphology persists at longer
wavelengths.  We used
  MERLIN\footnote{The Multi-Element Radio Linked Interferometer
  Network (MERLIN) is a UK national facility operated by the
  University of Manchester at Jodrell Bank Observatory on behalf of
  PPARC.}
for 50 hours in 4 sessions in 2003~February and April. We employed the
standard 1.65~GHz band, with a bandwidth of 15~MHz per polarization.
The data were calibrated using standard procedures in AIPS, and the
mapping was performed with CLEAN deconvolution and two iterations of
self-calibration (Figure~\ref{fig.J0743merlin}).  The 18~cm image
shows a string of compact components across the south side of the
lobe, and faint over-resolved emission in the rest of the lobe.  The
positions of the southern compact components agree well with the
position of the brightest emission on the south side of the VLA 6~cm
image. The north side of the lobe, however, does not show significant
compact emission at the VLA position of the B image.  Thus, the flux
ratio of the A and B images is significantly different at the two
wavelengths (A/B is 1.9 at 6~cm, but at least 3.0 at 18~cm), in
contrast to lensing theory and to our simple model (the 18~cm model is
shown in the lower panels of Figure~\ref{fig.J0743model}). This does
not strictly rule out the lensing hypothesis, since it could be viable
with some elaboration, such as preferential free-free absorption of
image B. However, in the absence of independent evidence for such an
effect, the lensing hypothesis is certainly less satisfying.

Another useful test is whether the mass of the galaxy as implied by
lensing theory is at least approximately equal to the mass of the
galaxy as inferred from its luminosity. The Tully-Fisher relation is
an empirical correlation between absolute magnitude and circular
velocity, and circular velocity can be taken as a proxy for mass.  The
following calculation is similar to that described more fully in
\cite{winn02c}. An isothermal sphere producing the 1\farcs8 image
separation between A and B would cause a flat rotation curve with
circular velocity 286~km~s$^{-1}$ (assuming the radio galaxy is at
redshift 1). Using the Tully-Fisher relation of \cite{sakai00a}, this
corresponds to an aperture absolute magnitude of the galaxy's disk of
$-24.6\pm0.4$ in rest-frame H-band.  We observed the H-band apparent
magnitude of the total disk to be $16.20\pm0.04$
(Table~\ref{table.optmodel}). After an aperture correction of
$2.5\log_{10}2$ \citep{aaronson82a, sakai00a}, a K-correction of
0.065~mag (as appropriate for an early type spiral with star formation
starting at redshift $\sim$1.5), and a 0.017~mag correction for
Galactic extinction \citep{schlegel98a}, the absolute magnitude is
estimated to be $-22.79\pm0.04$. This is $1.8\pm0.4$ magnitudes
(nearly 5$\sigma$) fainter than expected for a galaxy that is
sufficiently massive to be acting as a gravitational lens.  (This
result is relatively independent of source redshift: for $z=2$ the
discrepancy is still greater than 4$\sigma$, and for $z=0.3$ the
difference is over 7$\sigma$.)  Either the galaxy follows the
Tully-Fisher relation and does not have enough mass to produce images
A and B, or it has a much higher mass-to-light ratio than the local
galaxies on which the Tully-Fisher relation is based.

On balance, the case against lensing seems stronger than the case for
lensing.  The 6~cm morphology is certainly suggestive but the 18~cm
morphology does not obey the simplest lensing prediction, and the mass
of the galaxy as predicted by the Tully-Fisher relation is
insufficient to produce the 1\farcs8 image splitting. Additional
observations could confirm or deny the multiple-image interpretation
with more certainty.  For instance, detection of an optical core and
its photometric redshift could determine whether the radio lobe is
more distant than the spiral galaxy, and provide precise alignment for
better modeling. Higher resolution images at 6~cm would be sensitive
to fine structure in the lobe and could test whether the regions A and
B are as similar as they appear at lower resolution.

%%%%%%%%%%%%%%%%%%%%%%%%%%%%%%%%%%%%%%%%%%%%%%%%%%%%%%%%%%%%%%%%%%%%%%%
\section{Other Targets}
\label{other}

In this section we discuss other targets in the survey.  The status of
each of the 32 lens galaxy candidates is given in
Table~\ref{table.status}.  Here we describe some candidates in more
detail.  These have features that initially suggested lensing or
atypical lobes, prompting additional consideration or followup
observations.  None, however, turned out to be a promising lens
candidate worth further study.

\begin{figure}
\figurenum{1b}
\plotone{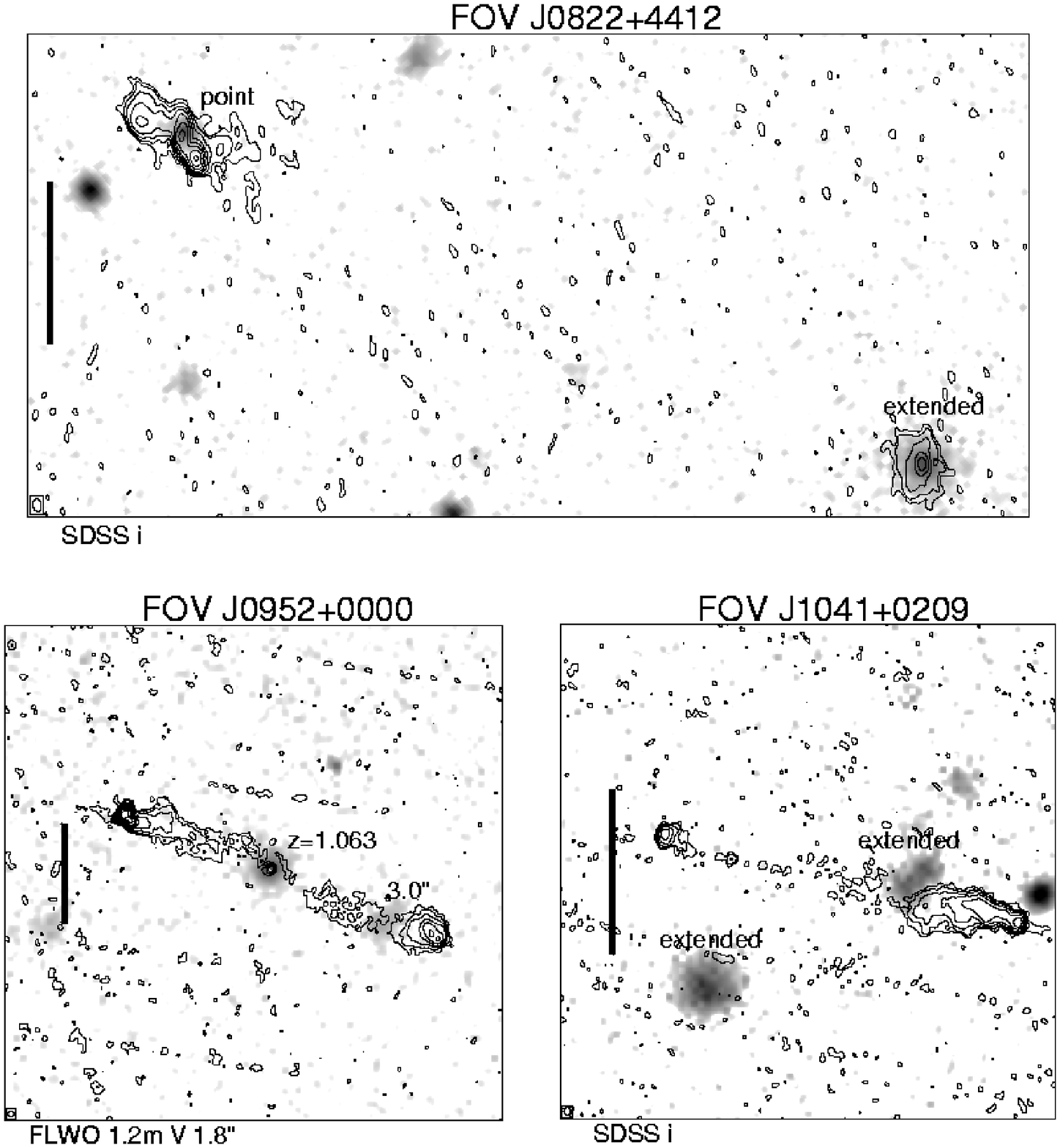}  %lenses2.eps
\caption{See Figure~\ref{fig.lenses1} caption. \label{fig.lenses2}}
\end{figure}

{\bf FOV~J0822+4412:} (Figure~\ref{fig.lenses2} There are two radio
components in the field, but it is unclear if they are physically
related to one another.  The middle of the eastern component is
associated with an unresolved SDSS optical object that may be the
active nucleus of the radio galaxy.  The western radio source is
extended, with an optical object having similar extent at the same
position; this radio emission is smooth with no sign of lensing or a
ring.  The western object could be a lobe of the eastern AGN with an
optical galaxy along the line of sight, or the western radio and
optical emission may originate from the same object, such as a
starburst galaxy.

{\bf FOV~J0952+0000}: (Figure~\ref{fig.lenses2}) The core of this
radio galaxy has an optical counterpart, giving a source redshift of
1.06 and precise optical-radio registration. The western jet has a
2\farcs5 gap at the position of an extended optical object at unknown
redshift.  If this galaxy is in the foreground, the gap in the jet may
be due in part to lensing (emission on either side of the gap could be
images of the same background source), or it could be an unlensed jet
with coincidental placement of the galaxy. The radio feature is quite
faint, and much deeper radio imaging would be required to begin to
separate the intrinsic shape of the jet from lensing effects.  

\begin{figure}
\figurenum{1c}
\plotone{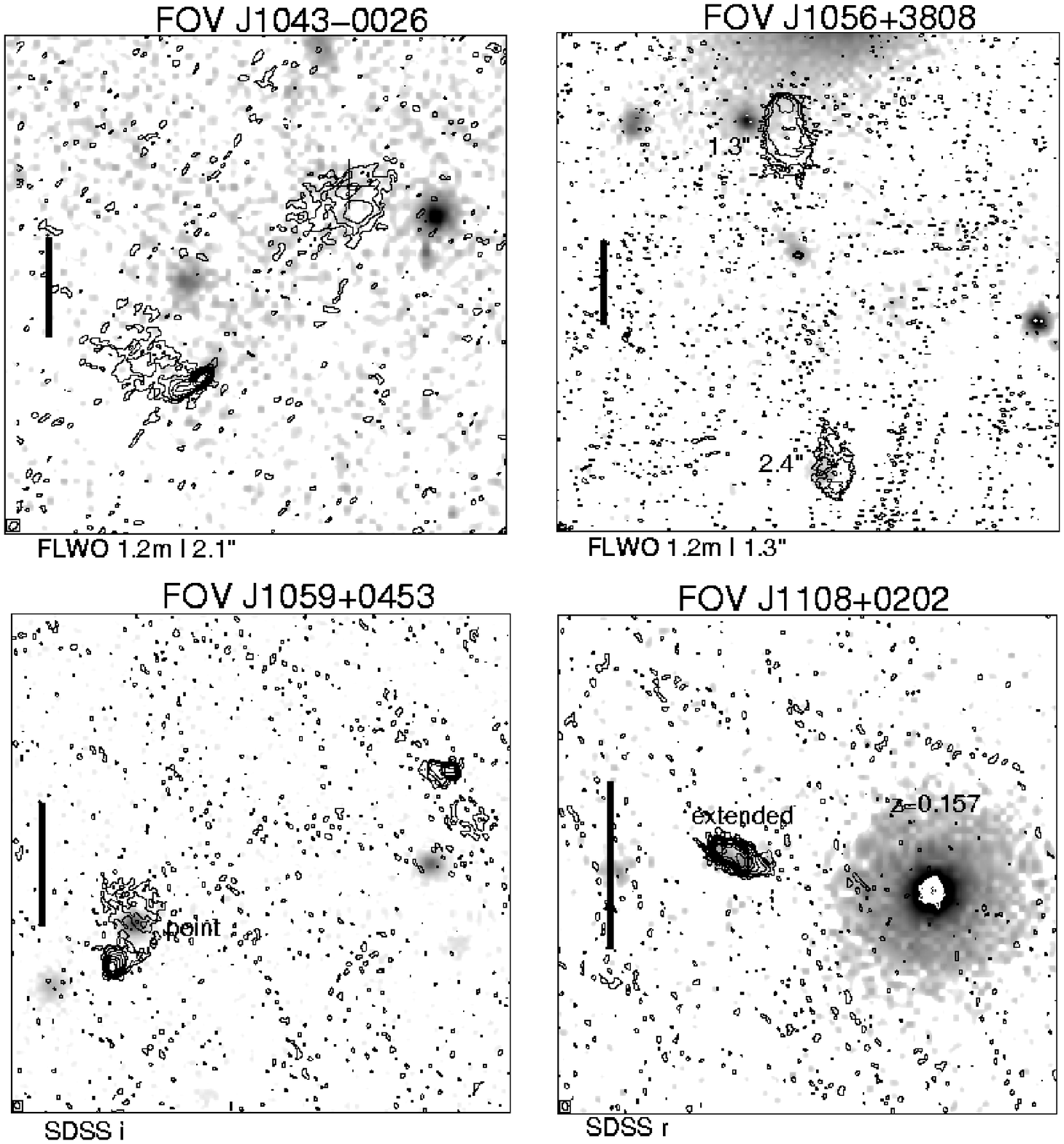} %lenses3.eps
\caption{See Figure~\ref{fig.lenses1} caption. \label{fig.lenses3}}
\end{figure}

{\bf FOV~J1108+0202:} (Figure~\ref{fig.lenses3}) The eastern radio
source has two closely spaced components on either side of an extended
optical object.  A polarization image (Figure~\ref{fig.J1108+.pol}),
however, shows that the two radio components have significantly
different fractional polarization and are very unlikely to be lensed
images of a single source.

\begin{figure}
\figurenum{1d}
\plotone{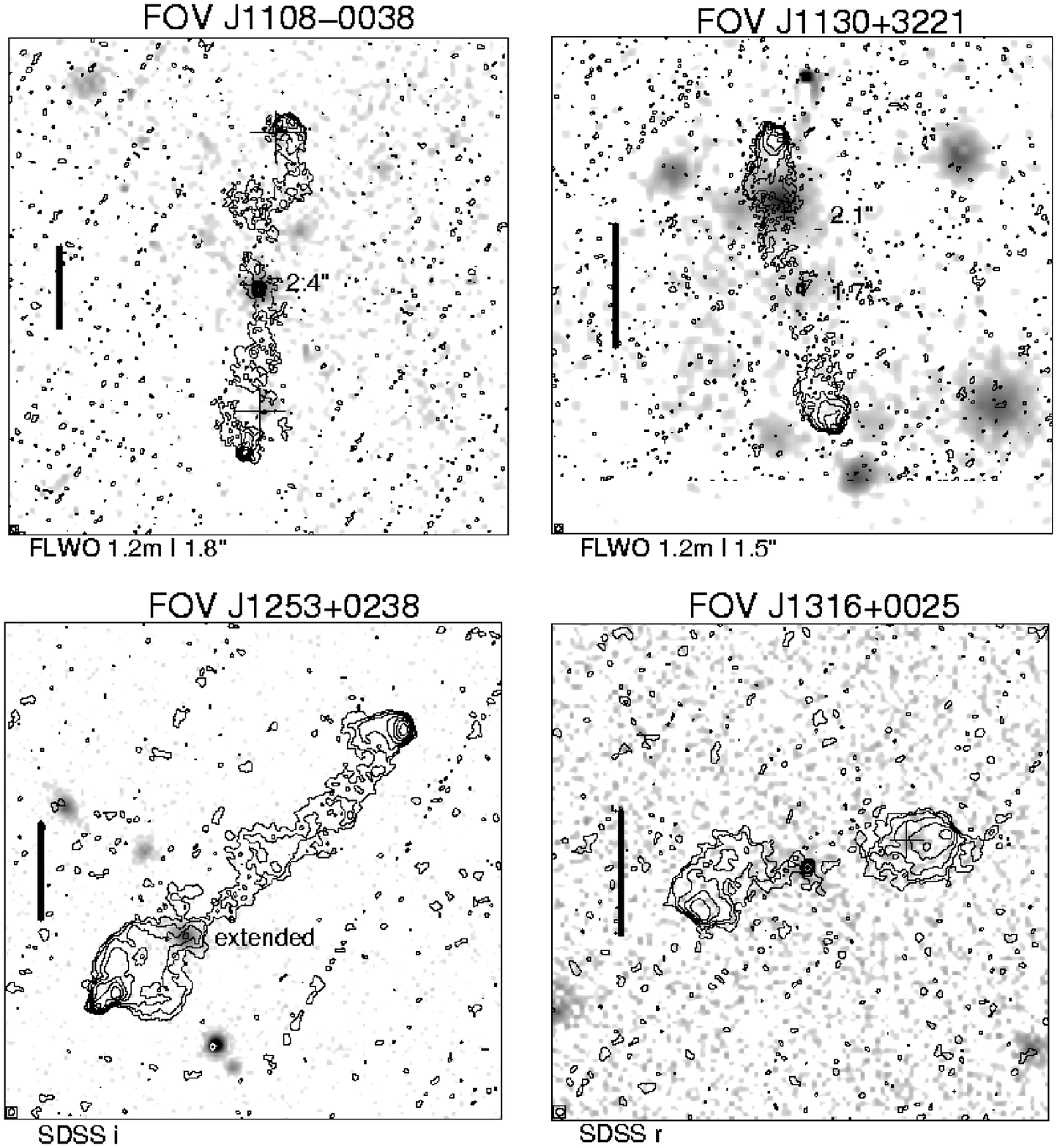} %lenses4.eps
\caption{See Figure~\ref{fig.lenses1} caption. \label{fig.lenses4}}
\end{figure}

{\bf FOV~J1108$-$0038 N}: (Figure~\ref{fig.lenses4}) The northern lobe
of this radio galaxy has an arc structure (about 3 arcseconds in
diameter), but the polarization map shows significantly different
polarization along the arc (Figure~\ref{fig.J1108-.pol}).  This
polarization is typical of radio lobes, with the electric field
perpendicular to the shock front.  Thus the arc is much more simply
explained as an intrinsic radio structure rather than a lensed lobe
with Faraday rotation along the arc.

{\bf FOV~J1253+0238:} (Figure~\ref{fig.lenses4}) The optical object is
extended (a galaxy), and coincident with a knot of radio emission.
This knot could be a hot spot in the jet, possibly lensed by a galaxy
in the foreground.  The optical-radio alignment suggests, however,
that the radio knot is the asymmetric core of the radio galaxy,
emitting jets to the northwest and southeast. The optical emission is
mostly likely from the AGN host galaxy at the same redshift.

{\bf FOV~J1316+0025:} (Figure~\ref{fig.lenses4}) A faint optical
object ($r=22.2$) is coincident with a bright smooth part of the
western radio lobe in the VLA image (Figure~1).  The core was detected
at both optical and radio wavelengths, allowing for precise
registration.  We observed this lobe at 18~cm using MERLIN on four
dates in 2003~April and May.  The data were calibrated using standard
techniques, and mapped using a maximum entropy deconvolution algorithm
(VTESS) and self-calibration (Figure~\ref{fig.J1316merlin}).  This
high resolution map shows that the optical counterpart is located far
from the brightest part of the lobe, and that there are no obvious
lensed structures around it.

{\bf FOV~J1508+0102:} The eastern lobe has a ring-like appearance in
the VLA 6~cm map (Figure~\ref{fig.J1508merlin}).  Initial optical
followup showed the optical counterpart to be extended and slightly
offset from the center of the radio ring.  The radio morphology was
confirmed in an 18~cm MERLIN observation in 2003~March, lasting 12
hours.  Comparison of the 6~cm and 18~cm maps shows that the spectral
index is similar for all the bright components of the ring, consistent
with a lensing interpretation.  Later optical images (with improved
seeing) showed, however, that the optical counterpart is two
unresolved objects, presumably a binary star.  It is possible that the
lobe is lensed but the lens galaxy is too faint to have been detected
in current CCD images, but it is also possible that it is an unlensed
lobe with unusual morphology.  Additional study is not warranted,
because even if it were lensed, the faintness of the lens galaxy would
make observations very difficult.

{\bf FOV~J1613+3724:} This object was originally selected because the
optical and radio catalog positions were coincident at the northwest
lobe.  Radio followup showed that the lobe was clearly south of the
optical catalog position (Figure~\ref{fig.lenses7}).  The ring-like
shape of the lobe prompted us to pursue optical followup anyway.
Unfortunately, no optical emission was detected within the radio ring
shape, and (similar to FOV~J1508+0102) further study is not warranted.

%%%%%%%%%%%%%%%%%%%%%%%%%%%%%%%%%%%%%%%%%%%%%%%%%%%%%%%%%%%%%%%%%%%%%%%
\section{Summary}
\label{conclusion}

The basic premise of our lensed lobe survey is that gravitational lens
candidates can be selected with relatively high efficiency from
existing wide-field radio and optical catalogs, by requiring radio
sources that appear to be lobes of radio galaxies to be nearly
coincident with optical sources that appear to be galaxies. We have
developed a selection algorithm for this purpose that is tailored to
the FIRST radio catalog, and either the APM or SDSS optical catalogs.
The superior depth and angular resolution of the SDSS catalog, in
particular, makes some of the optical follow-up observations
redundant, by greatly reducing the rate of false positive optical
detections and allowing a much stronger discrimination between stars
and galaxies.

In the previous round of this survey, using FIRST and APM,
\cite{lehar01a} identified one new lens and recovered one previously
known lens out of a sample of 33 candidate lensed lobes. In this
round, we selected an additional 92 candidate lensed lobes, mainly
from FIRST and APM but also using a subset of the SDSS catalog. After
follow-up observations with higher angular resolution at both radio
and optical wavelengths, we found 16 galaxies nearly coincident with
radio lobes, but these do not appear to be lensed. We found two
ring-shaped radio lobes, but detected no coincident optical galaxy
(FOV~J1508+0102 and J1613+3724). We found only one possible
gravitational lens (FOV~J0743+1553), but even in this case, the
lensing interpretation is problematic.  We must admit that this round
of the survey was not as successful in finding lenses as we had
expected, based on the first round. This warrants a reconsideration of
the basic premise underlying the survey.

Some of the challenges we experienced are based on an inherent
difficulty of any systematic study of radio lobes: their morphologies
are sometimes peculiar. This makes it difficult to establish with a
high level of confidence that a promising candidate is actually
gravitationally lensed, rather than simply exhibiting an intrinsically
lens-like morphology (an arc, a ring, or multiple blobs). An
additional challenge that is common to all radio lens surveys is that
the source redshift can be extremely difficult to measure. The radio
core must be detected and its optical counterpart must be bright
enough for spectroscopy.  This is not usually a problem in optical
lens surveys, for which the sources are (by design) bright quasars
with prominent emission lines.

Nevertheless, lensed lobes are worth pursuing because they are
plentiful and useful. They are plentiful because of the large angular
sizes of lobes relative to cores; they should constitute $\sim$75\% of
all radio lenses \citep{kochanek90a}. They are useful, at least in
principle, because observations of lensed extended sources provide
more information about the foreground mass distribution than lensed
point sources. To reap the benefits of lensed lobes, a survey strategy
needs to compensate for the difficulties discussed above by
efficiently producing large numbers of promising candidates, and in
particular, candidates that can be examined at high angular resolution
with a high signal-to-noise ratio.

Our pre-selection process did successfully find optical galaxies that
are nearly coincident with radio lobes (at least 17 out of 92
candidates). But, counter to our expectations, none of these 17
objects could be confidently interpreted as a case of gravitational
lensing. Why is this the case? In retrospect, we can offer some
plausible speculations. (1) The assumption that the galaxy is in the
foreground of the radio source may be incorrect. This assumption seems
reasonably well justified because the typical FIRST radio source has a
redshift of order unity, while the typical APM or SDSS galaxy has
redshift of $\sim$0.1. But there will surely be counter-examples. For
radio sources that are in clusters of galaxies, the optically detected
galaxy may be at nearly the same redshift as the radio source. This
may happen more often than one would naively expect, since a massive
galaxy is somewhat more likely to be a radio galaxy if it is in a
cluster, than if it is isolated. (2) The Einstein radius of the
optically detected galaxy may be too small to produce multiple images
or strong shearing of the background source, whether because of a
small galactic mass or an unfavorable combination of distances between
the lens and source.  For the typical masses and redshifts in the
parent catalogs of our sample, Einstein rings would be easily
detectable at the resolution of our VLA observations.  For at least
some of the actual targets we observed, the relative lens and source
distances would produce an Einstein ring too small for us to detect.
(3) The surface brightness distribution of the radio lobe may not have
enough contrast on arcsecond scales. If the surface brightness
distribution is too smooth on the angular size of the Einstein ring,
then gravitational lensing does not produce a detectable effect.

There is another respect in which our pre-selection process was
efficient: we were generally successful in identifying radio lobes
based on the limited morphological information of the comparatively
low-resolution FIRST catalog.  Our VLA observations of the 92 targets
showed that 83 of them were indeed lobes of radio
galaxies. Unfortunately none of these could be shown definitively to
be gravitationally lensed. As mentioned above, the lobe must vary in
brightness over angular sizes comparable to the Einstein radius, or
the Einstein ring will not appear. Yet the resolution of FIRST is
several times larger than the typical Einstein radius for our sample,
and hence FIRST cannot be used to pre-select lobes with sufficient
contrast. Conversely, some lobes that are nearly unresolved in FIRST
(and that might consequently be missed by our selection algorithm)
might have sufficient contrast on arcsecond scales.  Several
previously discovered lensed lobes (MG~0751+2716, MG~1131+0456, and
MG~1654+1346) appear unresolved or only slightly extended in FIRST
images.

Finally, even after drawing up a shortlist of candidates of definite
radio lobes with a definite candidate lens galaxy in the foreground,
the task still remains to obtain radio images with both high angular
resolution and high sensitivity. This must be done not only to confirm
that an object is lensed, but also to reap the benefits of the lensed
extended structures as modeling constraints, which is the primary
scientific motivation. Here the difficulty is that high angular
resolution generally means high-frequency observations (for a given
interferometer), and because the intrinsic radio spectra of radio
lobes generally decline sharply with frequency, the requirement of
high angular resolution is at odds with the requirement for high
sensitivity. We believe that a successful survey for lensed lobes will
require a new generation of radio interferometers, capable of rapid,
high-resolution imaging of steep-spectrum objects. With current
instruments, we have demonstrated that a careful pre-selection of
candidates reduces the total telescope time needed for the initial
phase of a lens survey, but the radio follow-up observations that are
needed to realize the benefits of lensed lobes are (at least in the
cases we have discovered in this round of the survey) prohibitively
expensive. Work is underway on the future instruments that would be
appropriate, such as the expanded VLA and e-MERLIN, and in the more
distant future, the Square Kilometer
Array\footnote{www.skatelescope.org} would be an outstanding tool to
find and characterize gravitationally lensed radio lobes.

%%%%%%%%%%%%%%%%%%%%%%%%%%%%%%%%%%%%%%%%%%%%%%%%%%%%%%%%%%%%%%%%%%%%%%%%%%

\acknowledgments

We are grateful to our referee for constructive criticism.  We thank
Ari Buchalter for assistance with FIRST catalog search software,
Richard McMahon for assistance with the APM-FIRST catalog, Ashish
Mahabel for assistance with the DPOSS catalog, David Rusin for
observations at Magellan, and Don Terndrup for observations at MDM.
D.B.H. was supported by a Cottrell College Science Award from Research
Corporation. For part of this work, J.N.W.\ was supported by an NSF
Astronomy and Astrophysics Postdoctoral Fellowship under award
AST-0104347, and subsequently by NASA through Hubble Fellowship grant
HST-HF-01180.02-A, awarded by the Space Telescope Science Institute,
which is operated by the Association of Universities for Research in
Astronomy, Inc., for NASA, under contract NAS~5-26555.  C.S.K.\ is
supported by NASA ATP grant NAG5-9265. This work is based on
observations made with the NASA/ESA Hubble Space Telescope, obtained
at the Space Telescope Science Institute, which is operated by AURA,
Inc. under NASA contract NAS5-26555; the observations are part of HST
program GO-9744.

%%%%%%%%%%%%%%%%%%%%%%%%%%%%%%%%%%%%%%%%%%%%%%%%%%%%%%%%%%%%%%%%%%%%%%
% Bibliography

%\clearpage
\bibliography{apj-jour,radio}
\bibliographystyle{apj}

%%%%%%%%%%%%%%%%%%%%%%%%%%%%%%%%%%%%%%%%%%%%%%%%%%%%%%%%%%%%%%%%%%%%
%\clearpage

\begin{figure}
\figurenum{1e}
\plotone{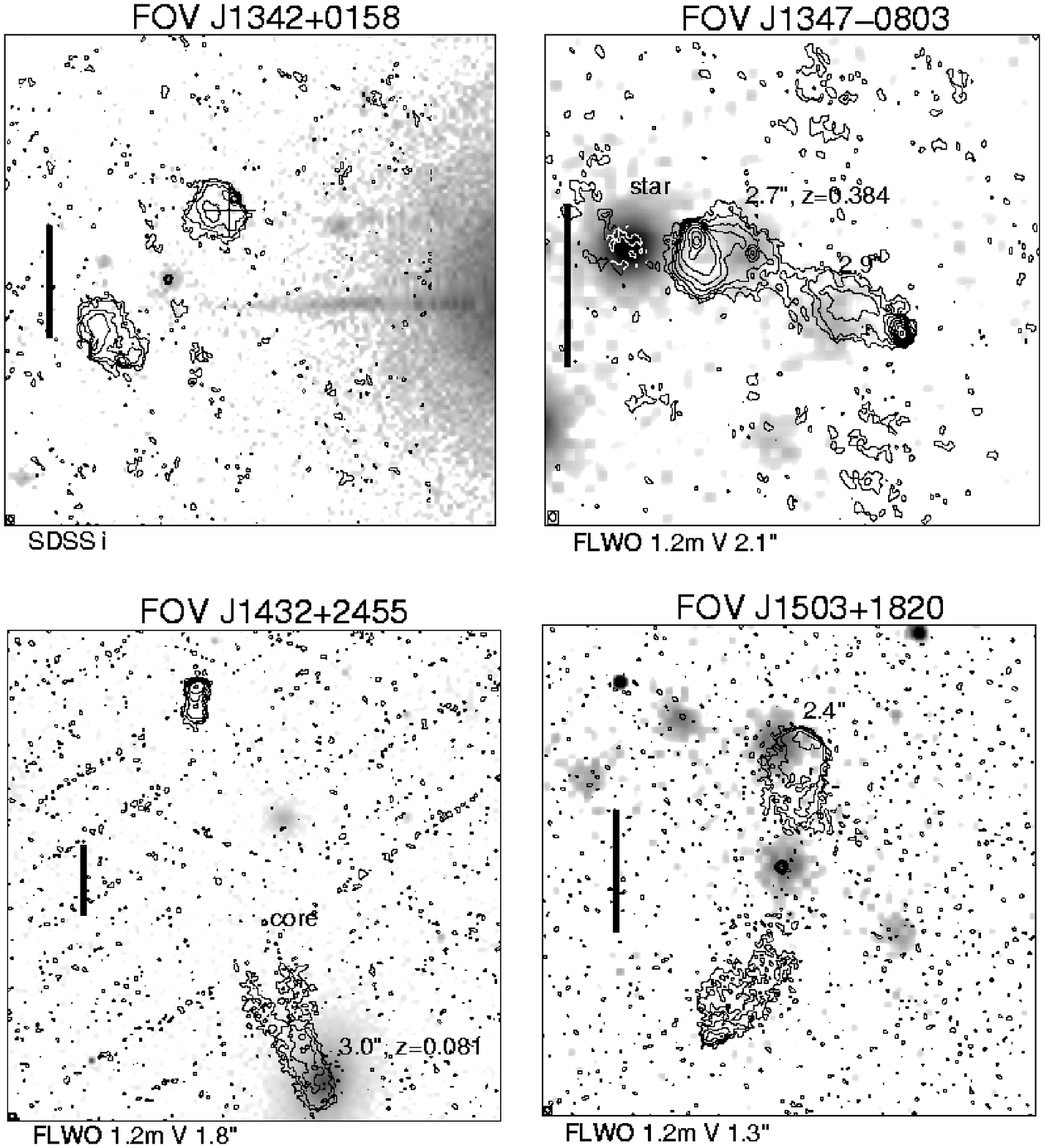}  %lenses5.eps
\caption{See Figure~\ref{fig.lenses1} caption. \label{fig.lenses5}}
\end{figure}

\begin{figure}
\figurenum{1f}
\plotone{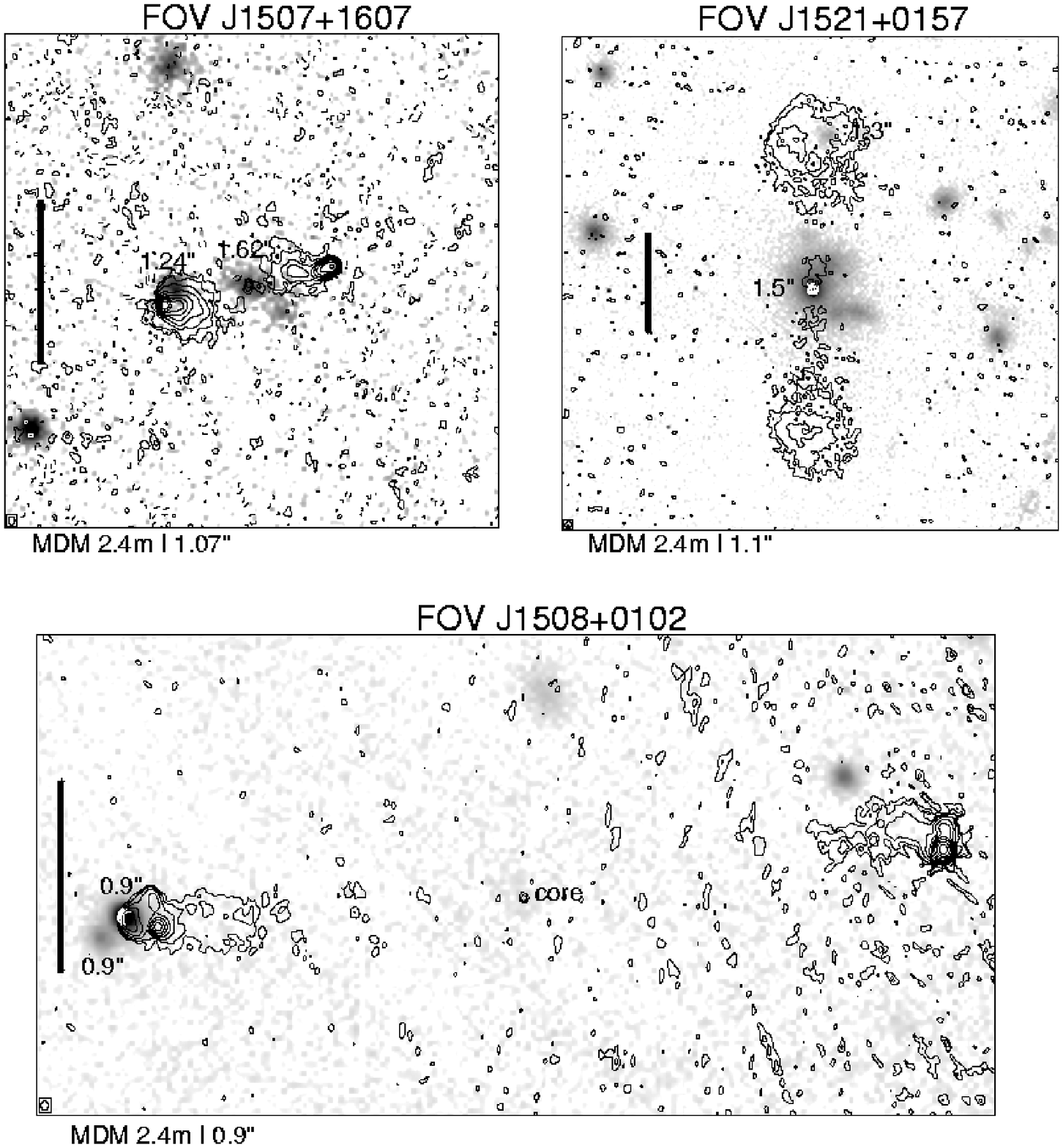}  %lenses6.eps
\caption{See Figure~\ref{fig.lenses1} caption. \label{fig.lenses6}}
\end{figure}

\begin{figure}
\figurenum{1g}
\plotone{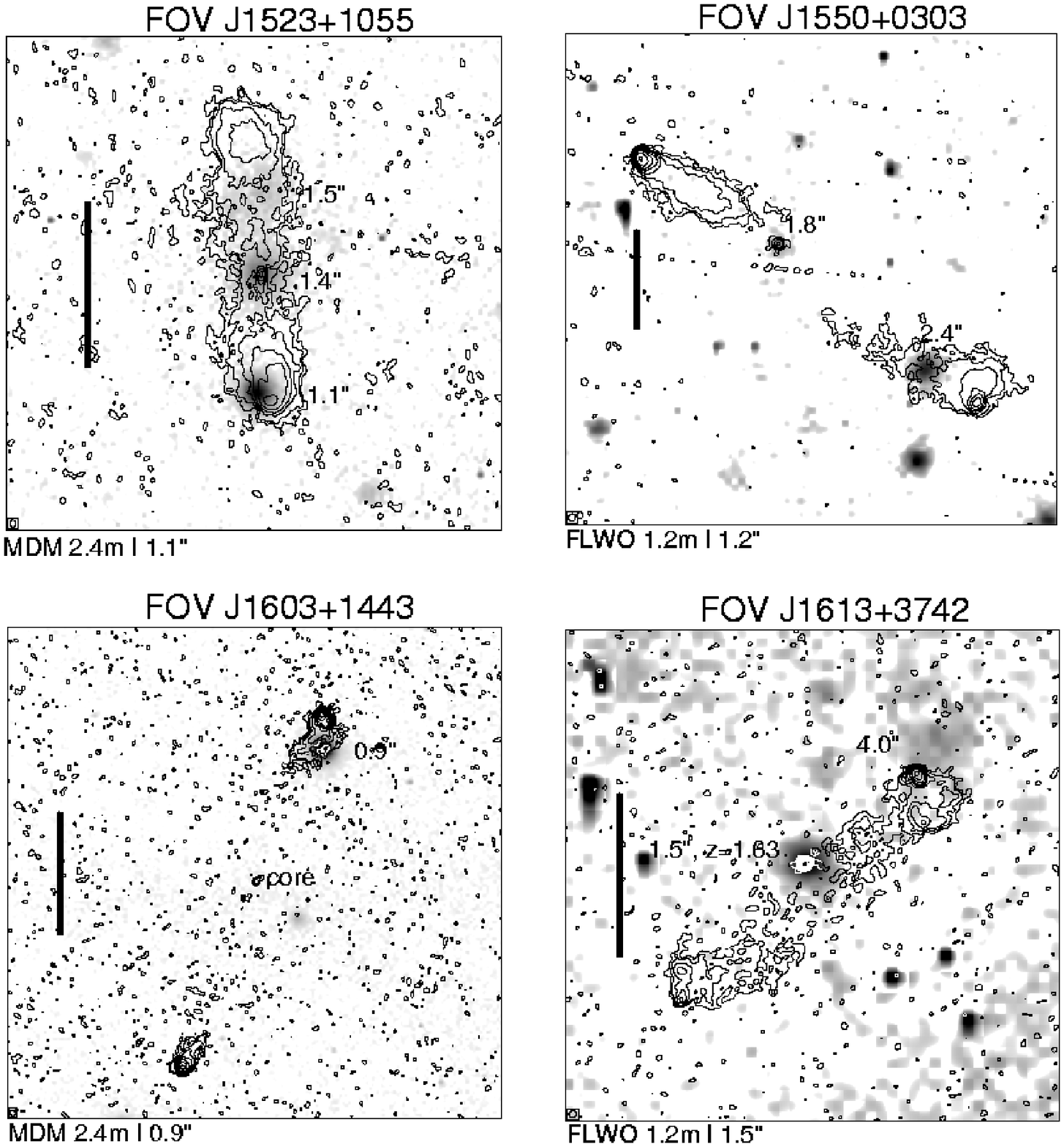}  %lenses7.eps
\caption{See Figure~\ref{fig.lenses1} caption. \label{fig.lenses7}}
\end{figure}

\begin{figure}
\figurenum{1h}
\plotone{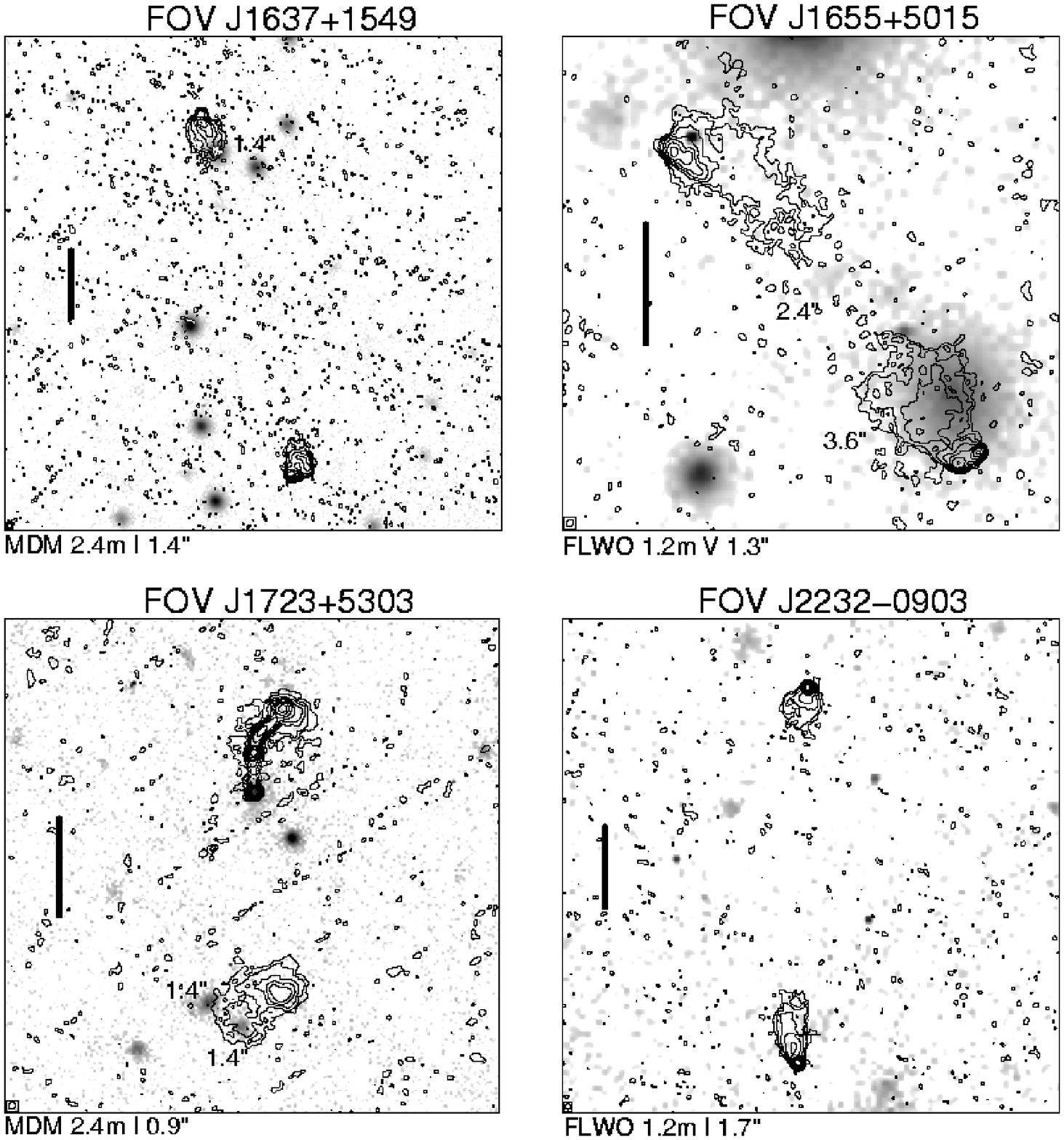} %lenses8.eps
\caption{See Figure~\ref{fig.lenses1} caption. \label{fig.lenses8}}
\end{figure}

\epsscale{0.5}
\begin{figure}
\figurenum{2}
\plotone{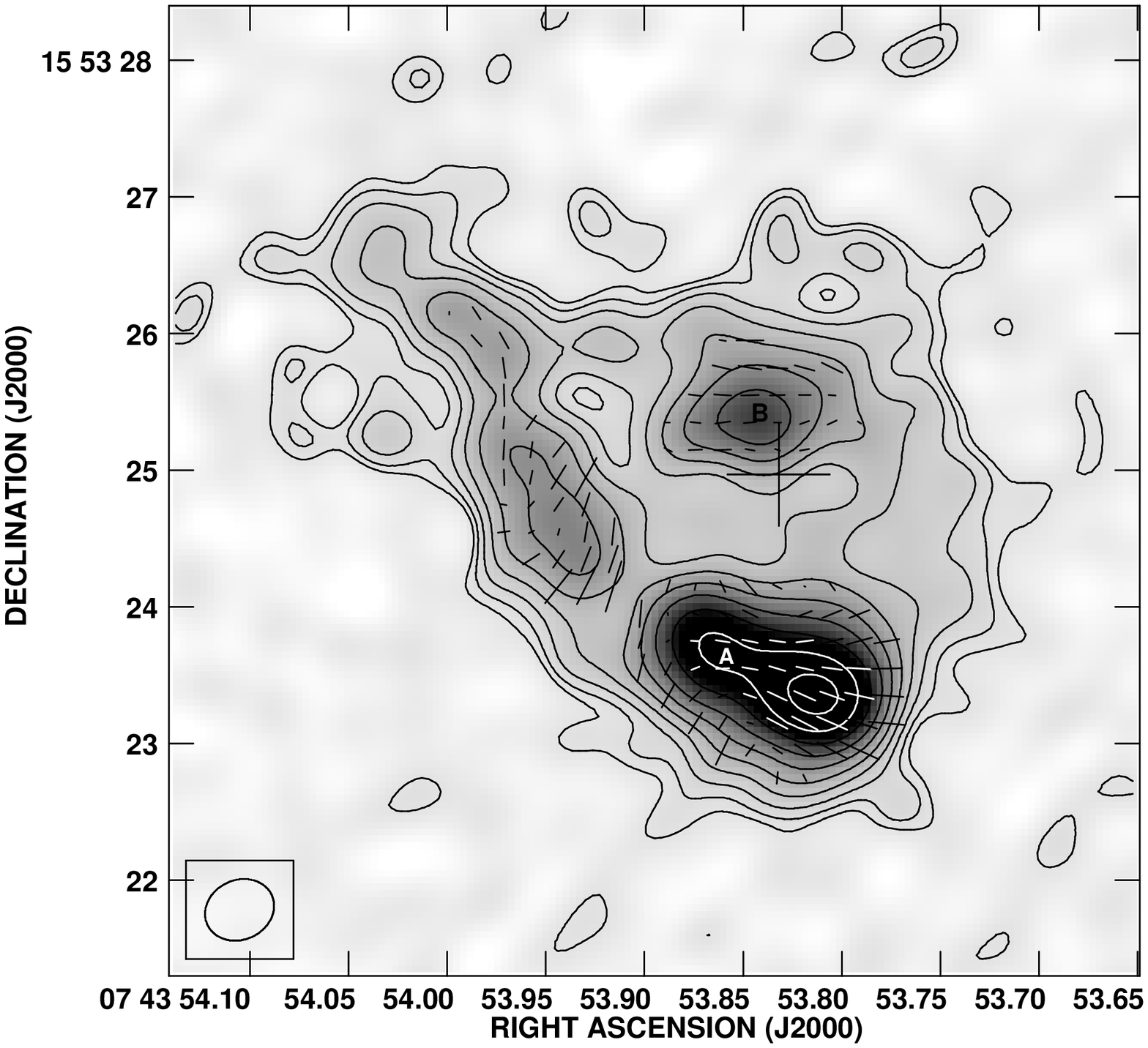} %J0743.VLA.label.eps}
\caption{VLA 6~cm map of gravitational lens FOV~J0743+1553.  Intensity
contours increase by factors of 1.5 from 1.5 times the off-source rms
noise of 50\mujy per beam.  The beam is shown in the lower left. The
fractional polarization vectors are parallel to the electric field and
scaled so that one spacing interval corresponds to 17\% polarization
per beam.  A cross indicates the measured position of the lens galaxy.
\label{fig.J0743VLA} 
}
\end{figure}

\epsscale{0.7}
\begin{figure}
\figurenum{3}
\plotone{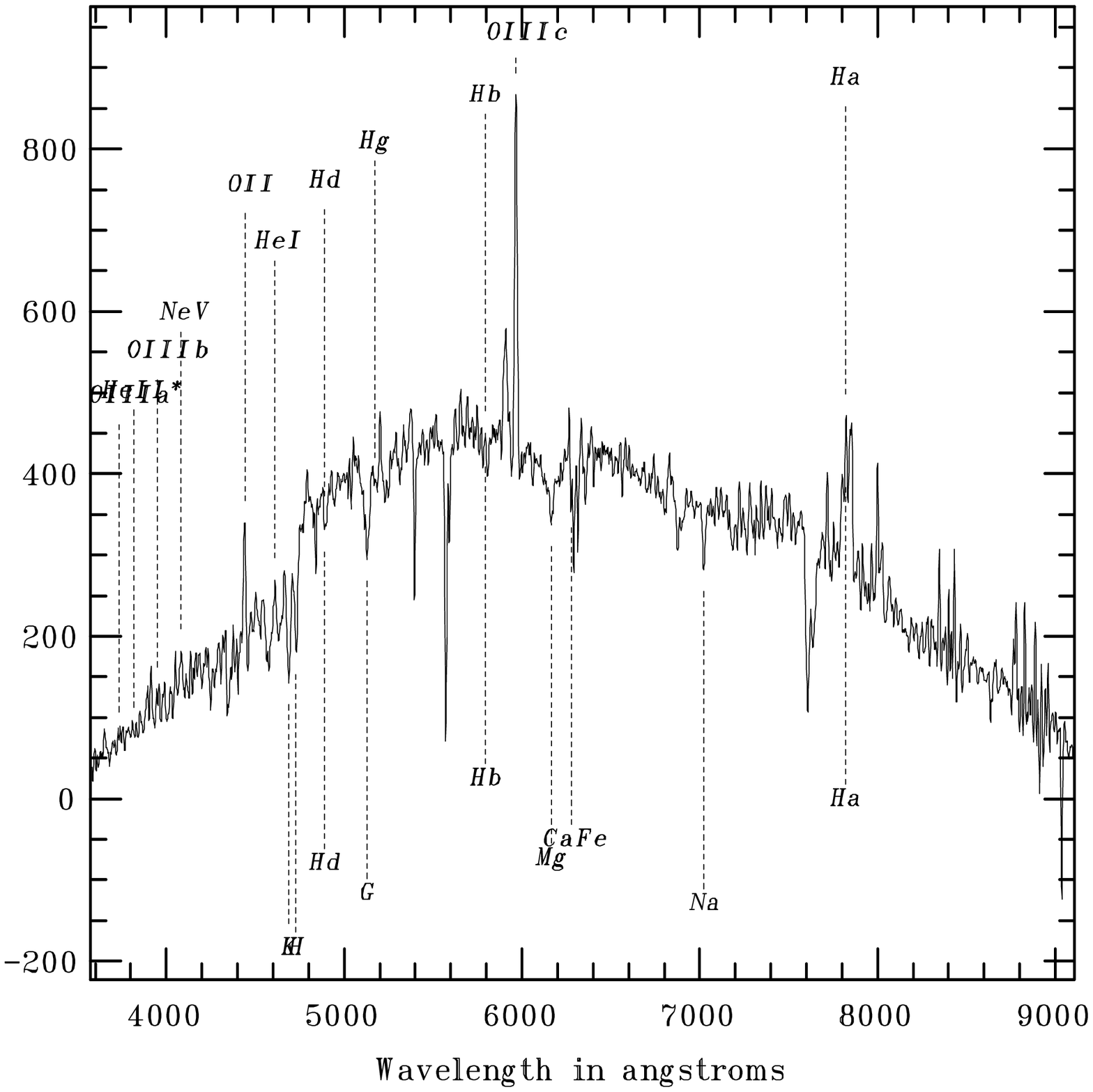} %J0743.spectrum.clean.eps}
\caption{Spectrum of lensing galaxy FOV~J0743+1553.  The data were
taken with the BlueChannel instrument on the 6.5~m MMT Observatory,
and show a lens redshift 0.1918.
\label{fig.J0743spect} 
}
\end{figure}

\epsscale{0.5}
\begin{figure}
\figurenum{4}
\plotone{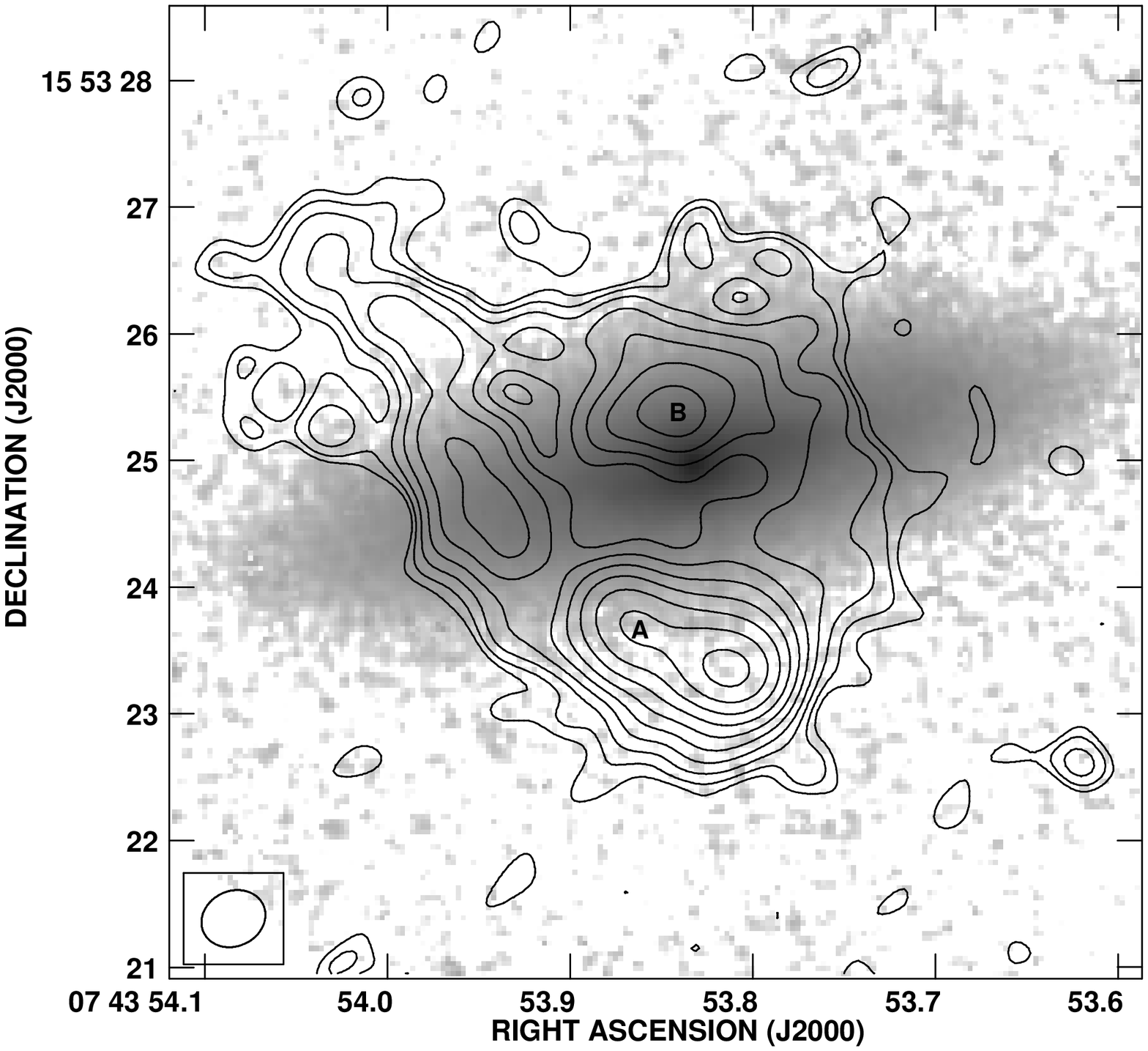}  %J0743.HSTH+VLA.label.eps}
\caption{Gravitational lens FOV~J0743+1553.  VLA 6~cm contours are
shown as in Figure~\ref{fig.J0743VLA}.  Hubble Space Telescope NICMOS 
data in the F160W filter is shown in grey with a logarithmic stretch.
\label{fig.J0743HSTH} 
}
\end{figure}

\epsscale{0.7}
\begin{figure}
\figurenum{5}
\plotone{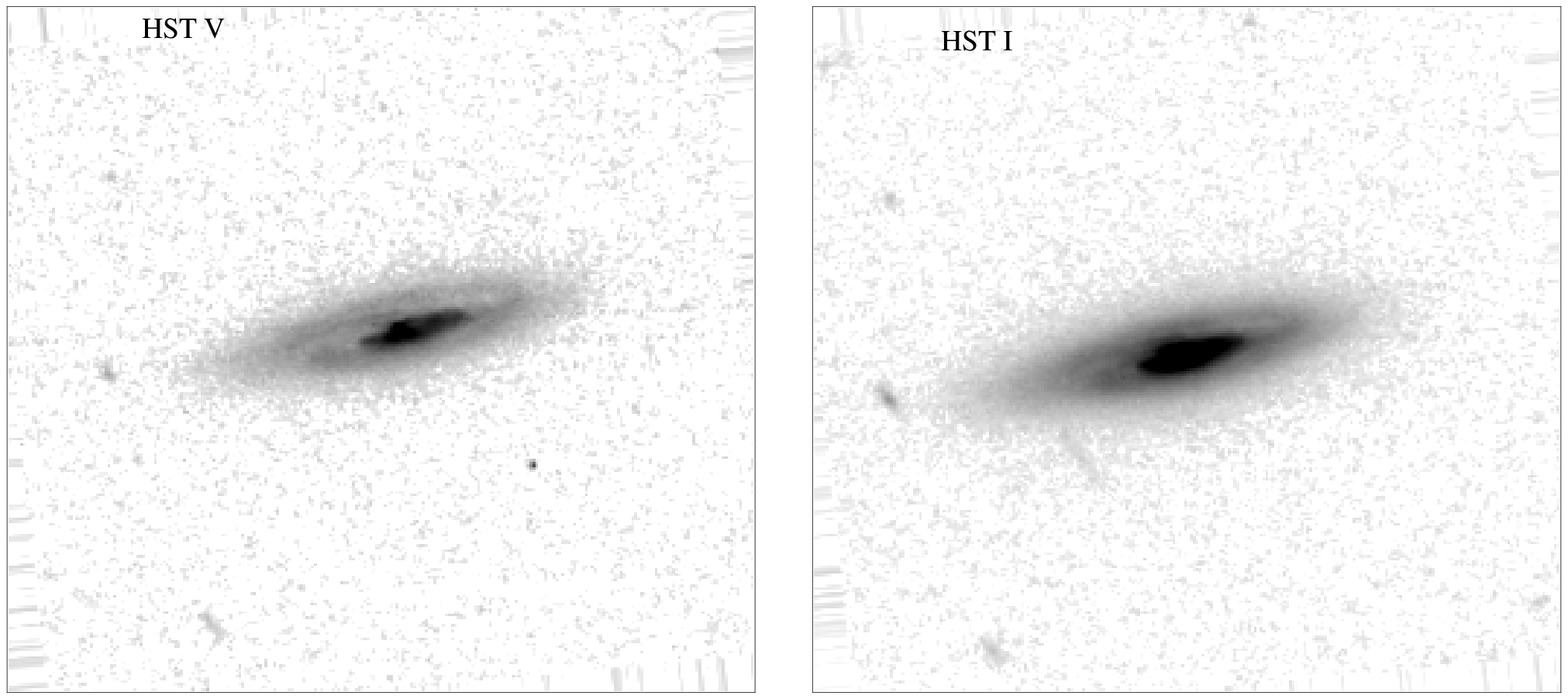} %J0743.HSTV+I.eps}
\caption{Gravitational lens FOV~J0743+1553.  Hubble Space Telescope
ACS data in V (F555W) and I (F814W) is shown with a square-root
stretch.  Maps are 13\arcsec$\times$13\arcsec\, with north up and east
to the left.
\label{fig.J0743HSTV+I} 
}
\end{figure}

\epsscale{0.5}
\begin{figure}
\figurenum{6}
\plotone{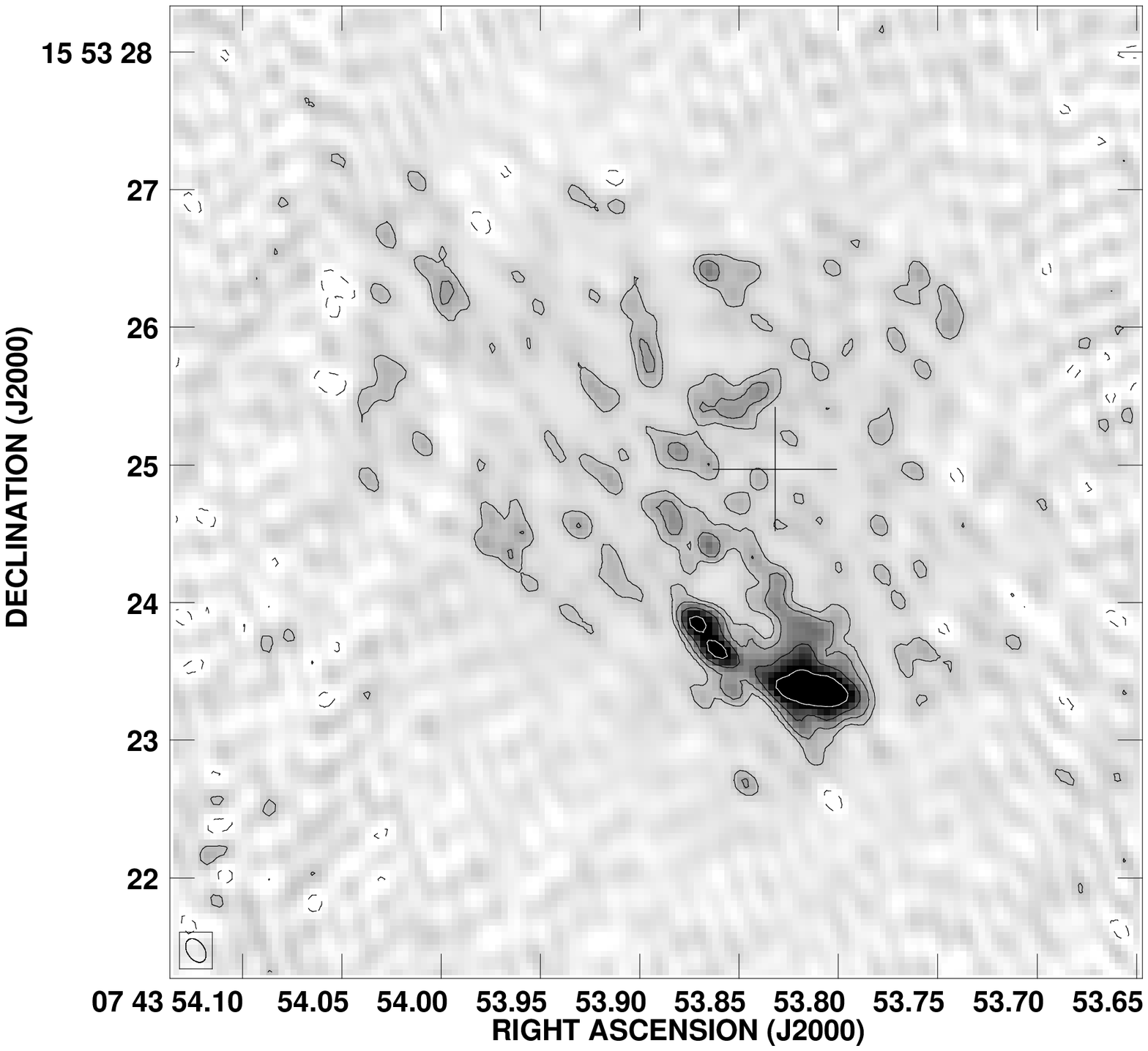} %J0743.merlin.ps}
\caption{MERLIN 18~cm map of FOV~J0743+1553.  Contours increase by
doubling from twice the off-source rms of 72\mujy per beam.  A cross
indicates the measured center of the lens galaxy.
\label{fig.J0743merlin}
}
\end{figure}

\epsscale{1.0}
\begin{figure}
\figurenum{7}
\plotone{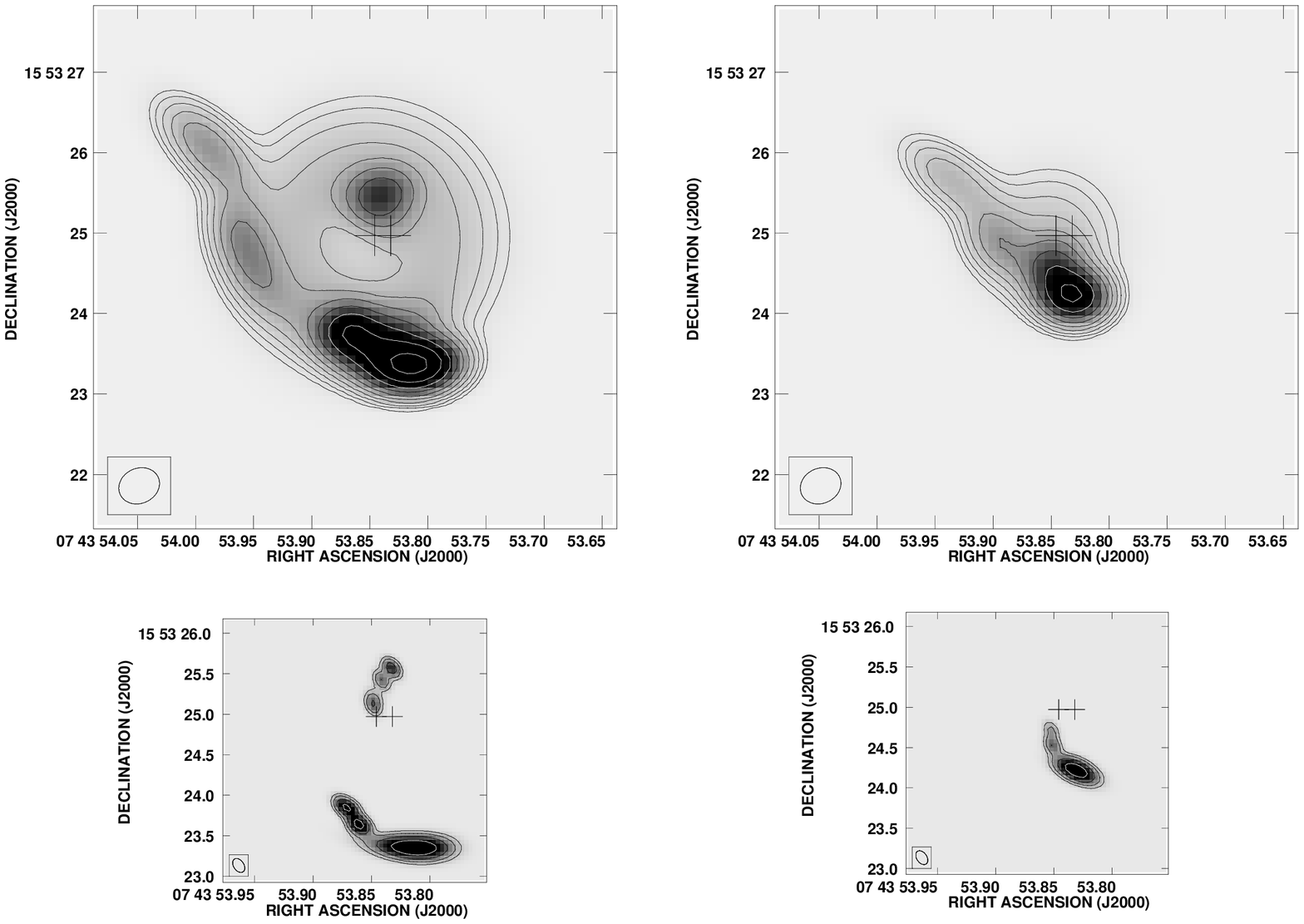}  %J0743.model.eps}
\caption{Model of FOV~J0743+1553.  Upper panels at 6~cm, lower panels
are 18~cm.  On the right, several Gaussian sources represent the
unlensed lobe.  On the left, a singular isothermal sphere is placed at
the location of the eastern cross, reproducing the lensed emission
(compare to Figures~\ref{fig.J0743VLA} and \ref{fig.J0743merlin}).
The western cross is the observed center of the lens galaxy.
\label{fig.J0743model} 
}
\end{figure}

\epsscale{1.0}
\begin{figure}
\figurenum{8}
\plotone{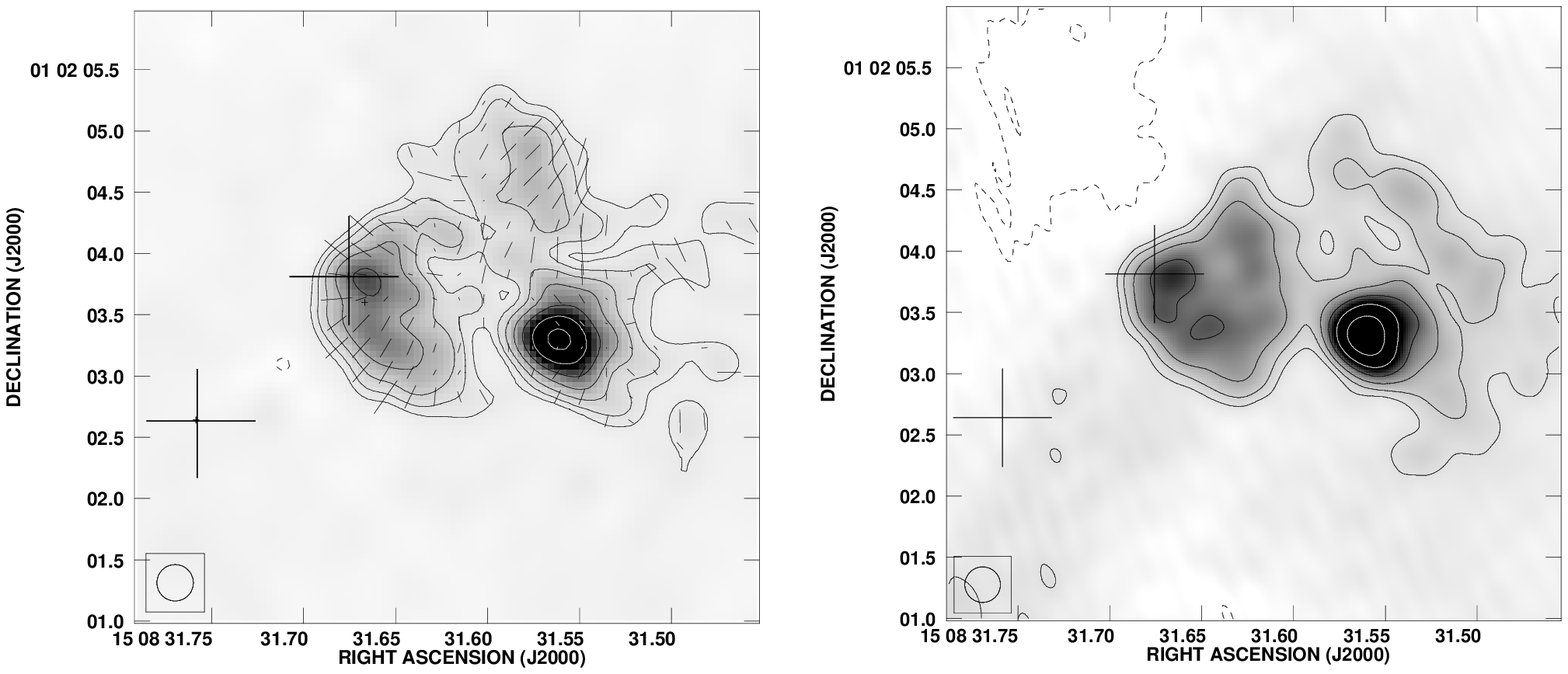} %J1508.vla+merlin.eps}
\caption{Radio maps of the east lobe of FOV~J1508+0102.  The 6~cm VLA
data is shown on the left, and the 18~cm MERLIN data is shown on the
right.  Both maps have been restored with a 0\farcs3 beam to allow
direct comparison.  Contours in both maps increase by doubling from
the off-source rms (74\mujy per beam in VLA map, 230\mujy per beam in MERLIN
map).  Crosses mark the location of the two stars in the MDM image.
In the 6~cm map, the fractional polarization vectors are parallel to
the electric field and scaled so that one spacing interval corresponds
to 25\% polarization per beam.
\label{fig.J1508merlin}
}
\end{figure}

\epsscale{0.7}
\begin{figure}
\figurenum{9}
\plotone{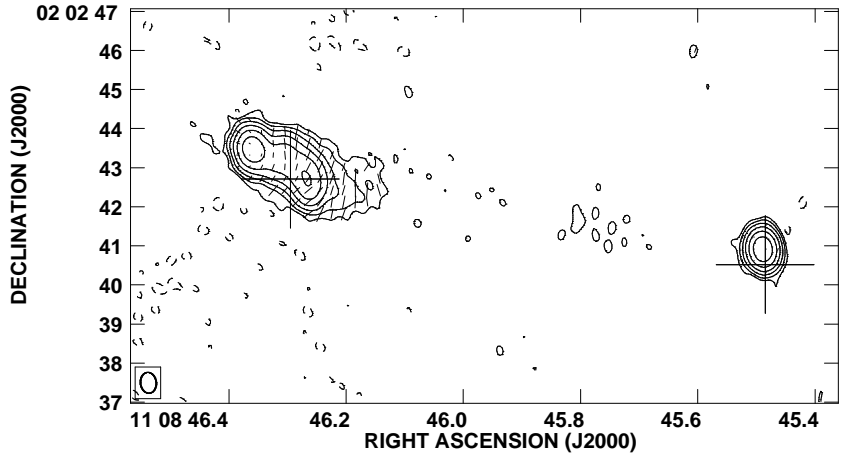}  %J1108+.pol.ps}
\caption{VLA 6~cm polarization map of FOV~J1108+0202.  Contours
increase by tripling from three times the off-source rms of 63\mujy.
The fractional polarization vectors are parallel to the electric field
and scaled so that one spacing interval corresponds to 11\%
polarization per beam.  
\label{fig.J1108+.pol} 
}
\end{figure}

\epsscale{0.5}
\begin{figure}
\figurenum{10}
\plotone{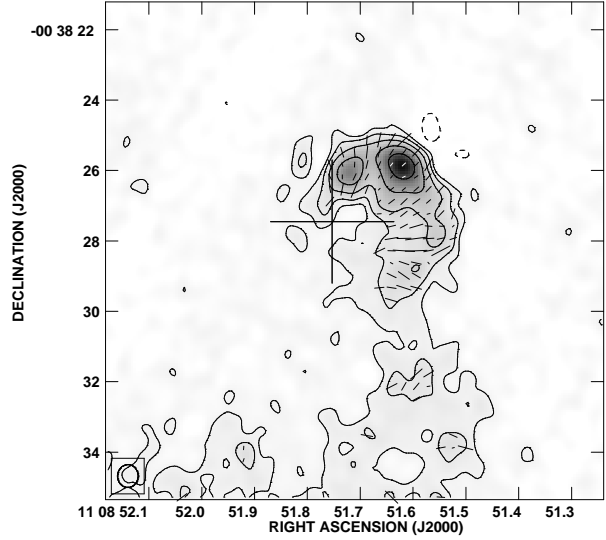}  %J1108-.pol.ps}
\caption{VLA 6~cm polarization map of the north lobe of
FOV~J1108-0038.  Contours increase by doubling from twice the
off-source rms of 40\mujy.  The fractional polarization vectors are
parallel to the electric field and scaled so that one spacing interval
corresponds to 20\% polarization per beam.
\label{fig.J1108-.pol} 
}
\end{figure}

\epsscale{0.7}
\begin{figure}
\figurenum{11}
\plotone{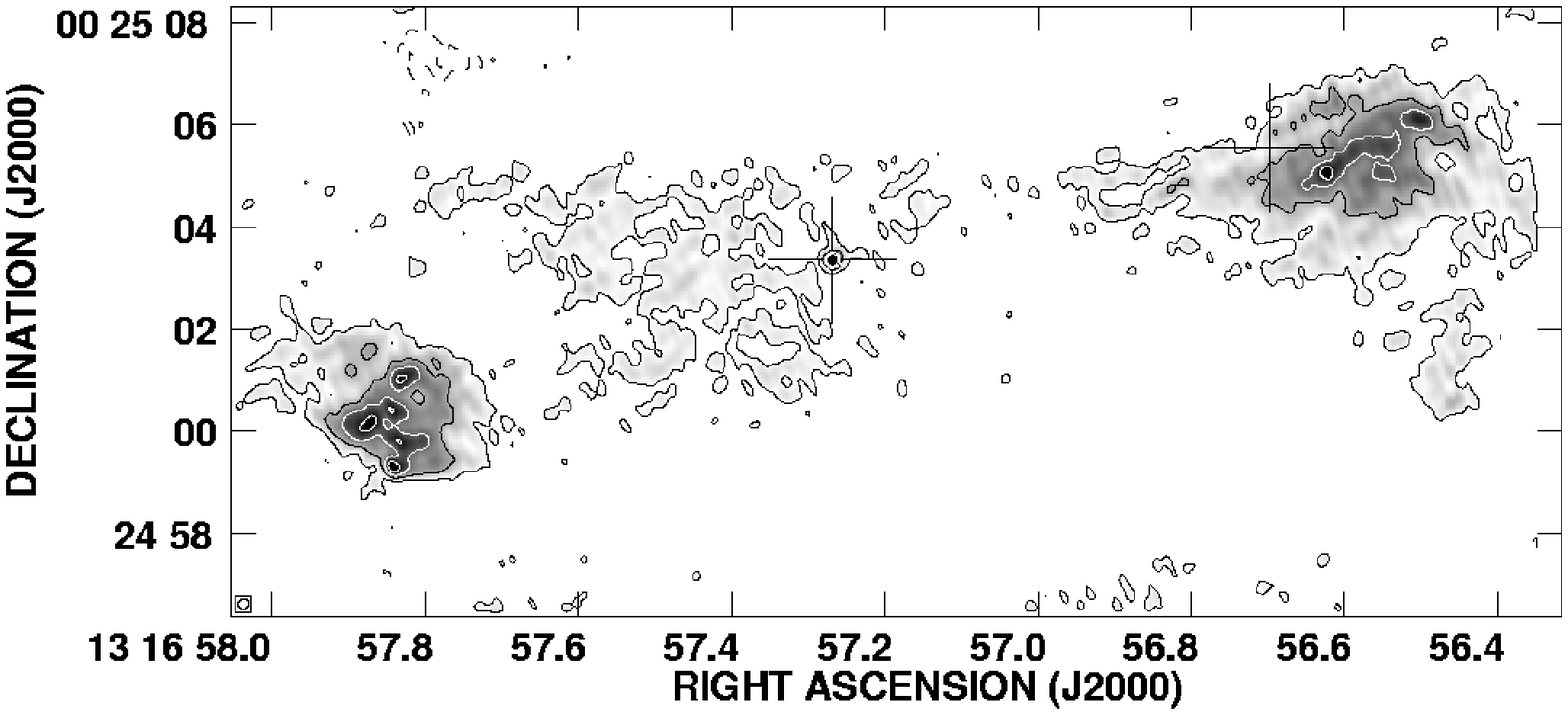} %J1316.merlin.ps}
\caption{MERLIN 18~cm map of FOV~J1316+0025.  Crosses indicate the
location of optical objects.  Contours increase linearly from twice
the off-source rms of 88\mujy.  The data were convolved with a
circular 0\farcs2 beam, shown in the lower right.
\label{fig.J1316merlin}
}
\end{figure}

\epsscale{1.0}
%%%%%%%%%%%%%%%%%%%%%%%%%%%%%%%%%%%%%%%%%%%%%%%%%%%%%%%%%%%%%%%
\clearpage
\begin{center}
\begin{deluxetable}{llccll}
\tabletypesize{\scriptsize}
\tablecaption{VLA observations of 92 targets
\label{table.vla} }
\tablehead{
\colhead{Name} & \colhead{Coordinates} & \colhead{Exp}   & \colhead{Wavelength} & \colhead{Date} & \colhead{Observer} \\ 
               & \colhead{(J2000)}     & \colhead{(min)} & \colhead{(cm)}      &                &       \\
}
\startdata
%VLA core / midpoint    exposure  Lambda obsdate  obs  verdict 
FOV J0040$-$0146 & 00:40:57.4 $-$01:46:18 & \nodata&\nodata& literature& VLA calibrator, point source \\
FOV J0040+0125   & 00:40:13.6   +01:25:46 & 18     & 6     & 2002Mar17 &              \\
FOV J0120+0004   & 01:20:53.9   +00:04:50 & 12     & 6     & 2002Mar17 &              \\
FOV J0121+0050   & 01:21:01.2   +00:51:01 & 18     & 6     & 2002Mar17 &              \\
FOV J0143$-$0118 & 01:43:16.7 $-$01:19:01 & 12     & 6     & 2002Mar17 &              \\
FOV J0202$-$0559 & 02:02:06.9 $-$05:59:00 & 18     & 6     & 2002Mar17 &              \\
FOV J0209$-$0736 & 02:09:31.4 $-$07:36:51 & 18     & 6     & 2002Mar17 &              \\
FOV J0233$-$0203 & 02:33:28.7 $-$02:02:56 & 18     & 6     & 2002Mar17 &              \\
FOV J0243$-$0102 & 02:43:49.4 $-$01:03:01 & 18     & 6     & 2002Mar17 &              \\
FOV J0735+4344   & 07:35:21.9   +43:44:21 & 10     & 6     & 1994Mar19 & C. Carilli   \\
FOV J0743+1553   & 07:43:54.2   +15:53:33 & 19     & 6     & 2002Mar15 &              \\
FOV J0750+2825   & 07:50:01.8   +28:25:10 & 12     & 6     & 2002Mar17 &              \\
FOV J0751+5301   & 07:51:58.9   +53:01:58 & 18     & 6     & 2002Mar17 &              \\
FOV J0804+1721   & 08:04:19.5   +17:21:33 & 19     & 6     & 2002Mar15 &              \\
FOV J0805+1243   & 08:05:23.4   +12:43:49 & 19     & 6     & 2002Mar15 &              \\
FOV J0811+2618   & 08:11:26.4   +26:18:15 & 11     & 4     & 1995Jul29 & K. Blundell  \\
FOV J0815+1741   & 08:15:34.4   +17:41:56 & 19     & 6     & 2002Mar15 &              \\
FOV J0822+4412   & 08:22:23.8   +44:12:37 & 19     & 6     & 2002Mar15 &              \\
FOV J0822+4614   & 08:22:08.3   +46:14:42 & 19     & 6     & 2002Mar15 &              \\
FOV J0912+5320   & 09:12:03.2   +53:20:28 & 19     & 6     & 2002Mar15 &              \\
FOV J0912$-$0708 & 09:12:00.5 $-$07:08:33 & 19     & 6     & 2002Mar15 &              \\
FOV J0914+1715   & 09:14:05.2   +17:15:54 & 12     & 6     & 2002Mar15 &              \\
FOV J0935+3633   & 09:35:30.4   +36:33:08 & 15     & 4     & 1990Jul19 & G. Taylor   \\ 
FOV J0939+0201   & 09:39:08.2   +02:01:02 & 19     & 6     & 2002Mar15 &              \\
FOV J0939+4016   & 09:39:20.4   +40:16:51 & 19     & 6     & 2002Mar15 &              \\
FOV J0945+1428   & 09:45:27.7   +14:28:17 & 21     & 6     & 2002Mar11 &              \\
FOV J0952+0000   & 09:52:45.7   +00:00:15 & 21     & 6     & 2002Mar11 &              \\
FOV J1004+1018   & 10:04:11.8   +10:18:42 & 21     & 6     & 2002Mar11 &              \\
FOV J1004$-$0715 & 10:04:23.6 $-$07:15:12 & 21     & 6     & 2002Mar11 &              \\
FOV J1040+2815   & 10:40:18.7   +28:15:19 & 19     & 6     & 2002Mar15 &              \\
FOV J1041+0209   & 10:41:23.4   +02:09:02 & 21     & 6     & 2002Mar09 &              \\
FOV J1043$-$0026 & 10:43:09.2 $-$00:26:29 & 21     & 6     & 2002Mar09 &              \\
FOV J1056+3808   & 10:56:12.4   +38:08:45 & 19     & 6     & 2002Mar15 &              \\
FOV J1059+0453   & 10:59:50.7   +04:53:57 & 21     & 6     & 2002Mar09 &              \\
FOV J1102+5550   & 11:02:26.4   +55:50:02 & 12     & 6     & 2002Mar15 &              \\
FOV J1103+5054   & 11:03:44.5   +50:54:02 & 19     & 6     & 2002Mar15 &              \\
FOV J1103$-$0512 & 11:03:33.7 $-$05:12:19 & 19     & 6     & 2002Mar15 &              \\
FOV J1106$-$0052 & 11:06:31.8 $-$00:52:52 & 21     & 6     & 2002Mar11 &              \\
FOV J1108+0202   & 11:08:46.0   +02:02:43 & 21     & 6     & 2002Mar11 &              \\
FOV J1108$-$0038 & 11:08:51.9 $-$00:38:46 & 21     & 6     & 2002Mar11 &              \\
FOV J1112$-$0025 & 11:12:31.1 $-$00:25:34 & 19     & 6     & 2002Mar15 &              \\
FOV J1117+1904   & 11:17:18.7   +19:04:24 & 12     & 6     & 2002Mar15 &              \\
FOV J1124+1142   & 11:24:41.4   +11:42:36 & 21     & 6     & 2002Mar11 &              \\
FOV J1129$-$0703 & 11:29:12.9 $-$07:03:20 & 21     & 6     & 2002Mar09 &              \\
FOV J1130+3221   & 11:30:33.8   +32:21:38 & 19     & 6     & 2002Mar15 &              \\
FOV J1134+0357   & 11:34:32.2   +03:57:46 & 21     & 6     & 2002Mar11 &              \\
FOV J1144+2958   & 11:44:21.3   +29:58:27 & 12     & 6     & 2002Mar15 &              \\
FOV J1145+4420   & 11:45:38.5   +44:20:22 & 12     & 6     & 2002Mar15 &              \\
FOV J1146+1805   & 11:46:32.1   +18:05:11 & 19     & 6     & 2002Mar15 &              \\
FOV J1204$-$0443 & 12:04:04.6 $-$04:43:53 & 21     & 6     & 2002Mar09 &              \\
FOV J1209$-$0257 & 12:09:06.9 $-$02:57:46 & 10     & 6     & 1982Jun11 & A. Downes    \\ 
FOV J1214$-$0100 & 12:14:48.9 $-$01:00:17 & 14     & 6     & 2002Mar11 &              \\
FOV J1215+0018   & 12:15:23.1   +00:18:55 & 12     & 6     & 2002Mar15 &              \\
FOV J1232$-$0224 & 12:32:00.3 $-$02:24:03 &\nodata &\nodata& literature& Kronberg et al. 1992 ApJ 387,528      \\
FOV J1253+0238   & 12:53:03.3   +02:38:25 & 13     & 6     & 2002Mar09 &              \\
FOV J1253+1139   & 12:53:13.4   +11:39:35 & 14     & 6     & 2002Mar11 &              \\
FOV J1300+3804   & 13:00:14.0   +38:04:30 & 19     & 6     & 2002Mar15 &              \\
FOV J1304+5324   & 13:04:44.2   +53:24:16 & 19     & 6     & 2002Mar15 &              \\
FOV J1316+0025   & 13:16:57.3   +00:25:03 & 21     & 6     & 2002Mar11 &              \\
FOV J1320+1743   & 13:20:21.2   +17:43:11 & 12     & 6     & 2002Mar15 &              \\
FOV J1331+5426   & 13:31:29.7   +54:26:30 & 19     & 6     & 2002Mar15 &              \\
FOV J1332+0101   & 13:32:16.5   +01:01:48 & 11     & 6     & 1982Jun12 & A. Downes   \\ 
FOV J1342+0158   & 13:42:48.4   +01:58:09 & 11     & 6     & 1982Jun12 & A. Downes   \\ 
FOV J1345$-$0252 & 13:45:52.2 $-$02:52:32 & 11     & 6     & 1982Jun12 & A. Downes   \\ 
FOV J1347$-$0000 & 13:47:45.1 $-$00:00:48 & 13     & 6     & 2002Mar09 &              \\
FOV J1347$-$0803 & 13:47:01.3 $-$08:03:25 & 13     & 6     & 2002Mar09 &              \\
FOV J1357+4807   & 13:57:30.6   +48:07:42 & 19     & 6     & 2002Mar15 &              \\
FOV J1405+2335   & 14:05:46.1   +23:35:53 & 19     & 6     & 2002Mar15 &              \\
FOV J1432+2455   & 14:32:42.2   +24:55:48 & 19     & 6     & 2002Mar15 &              \\
FOV J1444+1131   & 14:44:50.7   +11:31:56 & 21     & 6     & 2002Mar09 &              \\
FOV J1501+0752   & 15:01:57.5   +07:52:27 & 21     & 6     & 2002Mar11 &              \\
FOV J1503+1820   & 15:03:01.6   +18:20:31 & 19     & 6     & 2002Mar15 &              \\
FOV J1507+1607   & 15:07:09.1   +16:07:17 & 21     & 6     & 2002Mar11 &              \\
FOV J1508+0102   & 15:08:30.3   +01:02:06 & 21     & 6     & 2002Mar09 &              \\
FOV J1521+0157   & 15:21:27.9   +01:57:22 & 21     & 6     & 2002Mar09 &              \\
FOV J1523+1055   & 15:23:57.0   +10:55:43 & 14     & 6     & 2002Mar11 &              \\
FOV J1547+2748   & 15:47:21.0   +27:48:22 & 19     & 6     & 2002Mar15 &              \\
FOV J1550+0303   & 15:50:23.2   +03:03:47 & 21     & 6     & 2002Mar11 &              \\
FOV J1603+1443   & 16:03:05.5   +14:43:32 & 19     & 6     & 2002Mar15 &              \\
FOV J1605+0337   & 16:05:45.1   +03:37:17 & 19     & 6     & 2002Mar15 &              \\
FOV J1608+2848   & 16:08:11.3   +28:49:02 & 10     & 6     & 1981Mar17 & R. Sinha   \\ 
FOV J1613+3742   & 16:13:51.3   +37:42:58 & 19     & 6     & 2002Mar15 &              \\
FOV J1614+0405   & 16:14:14.5   +04:05:55 & 14     & 6     & 2002Mar11 &              \\
FOV J1637+1549   & 16:37:49.4   +15:49:22 & 19     & 6     & 2002Mar15 &              \\
FOV J1655+5015   & 16:55:22.1   +50:15:15 & 19     & 6     & 2002Mar15 &              \\
FOV J1710+4601   & 17:10:44.1   +46:01:29 &\nodata &\nodata& literature& 3C352      \\ 
FOV J1714+3001   & 17:14:10.1   +30:01:24 & 19     & 6     & 2002Mar15 &              \\
FOV J1723+5303   & 17:23:50.0   +53:03:02 & 19     & 6     & 2002Mar15 &              \\
FOV J1727+5230   & 17:27:17.7   +52:30:33 & 12     & 6     & 2002Mar15 &              \\
FOV J2232$-$0903 & 22:32:31.8 $-$09:03:37 & 18     & 6     & 2002Mar17 &              \\
FOV J2245$-$0204 & 22:45:43.4 $-$02:04:46 & 18     & 6     & 2002Mar17 &              \\
FOV J2333$-$1032 & 23:33:43.9 $-$10:32:20 & 18     & 6     & 2002Mar17 &              \\

\enddata
\tablecomments{Coordinates are of the center of the map (if shown in
Figure 1), the radio core (if present and unambiguous), or the
midpoint of the FIRST components (all others).}
\end{deluxetable}

\end{center}

%%%%%%%%%%%%%%%%%%%%%%%%%%%%%%%%%%%%%%%%%%%%%%%%%%%%%%%%%%%%%%%
%\clearpage
\begin{center}
 
\rotate
\tablewidth{0pt}
\begin{deluxetable}{ll lcc lcc cc }
\tabletypesize{\scriptsize}
\tablecaption{Optical data on 32 Targets
\label{table.opt} 
}
\tablehead{
               &                          &            &        &                       &\multicolumn{3}{c}{CCD Observations}                          &              &                    \\ \cline{6-8}  

\colhead{Name} &\colhead{Coordinates}     &            &        &                       &\colhead{Telescope,} &\colhead{Seeing}   &\colhead{Lens FWHM} &\multicolumn{2}{c}{Core detected?} \\ \cline{9-10}

               &\colhead{(J2000)}         & \multicolumn{3}{c}{Catalog Lens Magnitudes} &\colhead{Band}       & \colhead{(arcsec)} &\colhead{(arcsec)} &\colhead{VLA} &\colhead{CCD}       \\ \cline{3-5}
}
\startdata
%name                coord                       survey     R=       B           Tel     Band     see     lens    radcoreoptcore   redshift        % catalog. ccd
FOV J0121+0050     & 01 21 01.47   +00 50 47.6 & APM POSS & R=19.1 & B=21.9   & Magellan, i$'$  & 0.66   & 0.90   & yes  & yes    \\ 
FOV J0243$-$0102   & 02 43 49.12 $-$01 02 53.4 & SDSS EDR & r=22.6 & g=23.5   & Magellan, i$'$  & 1.8    & 2.2    & no   & no     \\ 
FOV J0743+1553     & 07 43 53.84   +15 53 24.9 & APM POSS & R=18.6 & B$>$21.5 & FLWO 1.2m, I    & 1.7    & 2.3    & yes  & no     \\ 
FOV J0822+4412 W   & 08 22 21.56   +44 12 24.2 & APM POSS & R=19.2 & B$>$21.5 & SDSS DR1,  i    &\nodata &extended& yes  & yes    \\ 
FOV J0939+4016     & 09 39 20.63   +40 17 02.8 & APM POSS & R=19.1 & B=21.7   & FLWO 1.2m, I    & 1.6    & 1.8    & yes  & no     \\
FOV J0952+0000     & 09 52 44.81   +00 00 10.0 & APM UKST & R=19.6 & B=20.7   & FLWO 1.2m, I    & 1.8    & 3.0    & yes  & yes    \\ 
FOV J1041+0209     & 10 41 22.90   +02 09 02.2 & APM UKST & R=19.3 & B=20.4   & SDSS DR1,  i    &\nodata &extended& yes? & no     \\ 
FOV J1043$-$0026   & 10 43 08.58 $-$00 26 19.0 & SDSS EDR & r=23.0 & g=23.4   & FLWO 1.2m, I    & 2.1 & detected? & yes  & ?      \\ 
FOV J1056+3808     & 10 56 12.24   +38 08 21.5 & APM POSS & R$>$20 & B=22.1   & FLWO 1.2m, I    & 1.3    & 2.4    & yes  & yes    \\
FOV J1059+0453     & 10 59 51.38   +04 53 52.0 & APM POSS & R=17.9 & B$>$21.5 & SDSS DR1,  i    &\nodata & point  & no   & no     \\
FOV J1108+0202     & 11 08 46.28   +02 02 42.7 & APM UKST & R$>$21 & B=21.0   & SDSS DR1,  r    &\nodata &extended& yes  & yes    \\
FOV J1108$-$0038 N & 11 08 51.75 $-$00 38 27.5 & SDSS EDR & r=22.5 & g=23.9   & FLWO 1.2m, I    & 1.8 & detected? & yes  & yes    \\ 
FOV J1108$-$0038 S & 11 08 51.86 $-$00 39 01.4 & SDSS EDR & r=21.4 & g=25.0   &                 &     & detected  &      &        \\ 
FOV J1130+3221     & 11 30 33.94   +32 21 44.9 & APM POSS & R=18.6 & B$>$20.6 & FLWO 1.2m, I    & 1.5    & 2.1    & yes  & yes    \\
FOV J1253+0238     & 12 53 03.72   +02 38 18.3 & APM POSS & R=18.3 & B=21.2   & SDSS DR1,  i    &\nodata &extended& yes  & yes    \\
FOV J1316+0025     & 13 16 56.70   +00 25 05.6 & SDSS EDR & r=22.2 & g=22.4   & SDSS DR1,  i    &\nodata &detected& yes  & yes    \\ 
FOV J1342+0158     & 13 42 48.00   +01 58 15.5 & APM UKST & R$>$21 & B=21.7   & SDSS DR1,  i    &\nodata &not detected& yes & yes \\
FOV J1347$-$0803   & 13 47 01.47 $-$08 03 23.3 & APM UKST & R=19.9 & B=21.6   & FLWO 1.2m, I    & 2.1    & 2.9    & yes  & yes    \\
FOV J1432+2455     & 14 32 41.46   +24 55 18.2 & APM POSS & R=15.1 & B=16.9   & FLWO 1.2m, I    & 1.8    & 3.0    & yes  & no     \\
FOV J1503+1820     & 15 03 01.67   +18 20 41.4 & APM POSS & R=19.0 & B=20.6   & FLWO 1.2m, I    & 1.3    & 2.4    & yes  & yes    \\
FOV J1507+1607     & 15 07 09.42   +16 07 16.3 & DPOSS    & i=20.9 & g=20.9   & MDM 2.4m,  I    & 1.07   & 1.24   & yes  & yes    \\
FOV J1508+0102     & 15 08 31.68   +01 02 03.8 & APM UKST & R=19.0 & B=21.0   & MDM 2.4m,  I    & 0.9    & 0.9    & yes  & ?      \\ 
FOV J1521+0157     & 15 21 27.79   +01 57 36.8 & APM UKST & R$>$21 & B=23.6   & MDM 2.4m,  I    & 1.1    & 1.3    & yes  & yes    \\ 
FOV J1523+1055 N   & 15 23 56.96   +10 55 48.5 & APM POSS & R$>$20 & B=21.2   & MDM 2.4m,  I    & 1.1    & 1.5    & yes  & yes    \\ 
FOV J1523+1055 S   & 15 23 56.99   +10 55 36.0 & DPOSS    & r=19.8 & g=21.5   &                 &        & 1.1    &      &        \\ 
FOV J1550+0303     & 15 50 22.41   +03 03 37.7 & APM POSS & R=19.4 & B$>$21.5 & FLWO 1.2m, I    & 1.2    & 2.4    & yes  & yes    \\ 
FOV J1603+1443     & 16 03 05.11   +14 43 42.3 & APM POSS & R=17.5 & B=20.6   & MDM 2.4m,  I    & 0.9    & 0.9    & yes  & no     \\ 
FOV J1613+3742     & 16 13 50.7\phno +37 43 05.8 & APM POSS & R$>$20 & B=21.2 & FLWO 1.2m, I    & 1.5    & 4.0    & yes  & yes    \\
FOV J1637+1549     & 16 37 49.76   +15 49 41.0 & APM POSS & R=18.5 & B$>$21.5 & MDM 2.4m,  I    & 1.4    & 1.4    & no   & no     \\
FOV J1655+5015     & 16 55 20.91   +50 15 05.9 & APM POSS & R=17.5 & B=18.9   & FLWO 1.2m, I    & 1.3    & 3.6    & no   & ?      \\
FOV J1723+5303     & 17 23 50.14   +53 02 45.5 & SDSS EDR & r=22.0 & g=25.1   & MDM 2.4m,  I    & 0.9    & 1.4    & yes  & yes    \\ 
FOV J2232$-$0903   & 22 32 31.82 $-$09 03 57.7 & APM UKST & R$>$21 & B=22.3   & FLWO 1.2m, I    & 1.7    &detected? & no   & no     \\
\enddata

\tablecomments{Objects
FOV~J1108$-$0038 and FOV~J1523+1055 each have two potential lens galaxies.
Objects FOV~J1507+1607 and FOV~J1523+1055 were selected based on an
APM catalog entry, but some components in the image didn't appear in
the APM catalog, so DPOSS data were used (DPOSS is the Digitized
Palomar Observatory Sky Survey,
http://taltos.pha.jhu.edu/$\sim$rrg/science/dposs).  
}
\end{deluxetable}

\end{center}

%%%%%%%%%%%%%%%%%%%%%%%%%%%%%%%%%%%%%%%%%%%%%%%%%%%%%%%%%%%%%%%
%\clearpage
\begin{center}
\tabletypesize{\footnotesize} 
%\rotate
\tablewidth{0pt}
\begin{deluxetable}{lll}
\tablecaption{Status of 32 Targets
\label{table.status} 
}
\tablehead{
\colhead{Name}     & Optical status          & Radio status
}
\startdata
FOV J0121+0050     & aligned galaxy             & lobe, $z=0.237$              \\ 
FOV J0243$-$0102   & aligned galaxy             & lobe                      \\ 
FOV J0743+1553     & aligned galaxy, $z=0.1918$ & lobe with possible lensed features \\ 
FOV J0822+4412 W   & starburst galaxy?          & starburst galaxy?         \\ 
FOV J0939+4016     & aligned galaxy, $z=0.186$  & lobe                      \\
FOV J0952+0000     & aligned galaxy             & lobe, $z=1.063$ \\ 
FOV J1041+0209     & not aligned                & lobe                      \\ 
FOV J1043$-$0026   & extent not measured        & lobe                      \\ 
FOV J1056+3808     & aligned galaxy             & lobe                      \\
FOV J1059+0453     & star                       & lobe                      \\
FOV J1108+0202     & aligned galaxy             & lobe, $z=0.157$ \\
FOV J1108$-$0038 N & extent not measured        & lobe \\
FOV J1108$-$0038 S & extent not measured        & lobe \\
FOV J1130+3221     & aligned galaxy             & lobe                      \\
FOV J1253+0238     & aligned galaxy             & core                      \\
FOV J1316+0025     & extent not measured        & lobe                      \\ 
FOV J1342+0158     & spurious catalog entry     & lobe                      \\
FOV J1347$-$0803   & aligned galaxy             & lobe, $z=0.384$             \\
FOV J1432+2455     & aligned galaxy, $z=0.081$  & lobe             \\
FOV J1503+1820     & aligned galaxy             & lobe                      \\
FOV J1507+1607     & aligned galaxy             & lobe                      \\
FOV J1508+0102     & double star                & lobe with possible lensed features \\  
FOV J1521+0157     & aligned galaxy             & lobe                      \\ 
FOV J1523+1055 N   & aligned galaxy             & lobe                      \\ 
FOV J1523+1055 S   & star                       & lobe                      \\ 
FOV J1550+0303     & aligned galaxy             & lobe                      \\ 
FOV J1603+1443     & star                       & lobe                      \\ 
FOV J1613+3742     & not aligned                & lobe with possible lensed features, $z=1.63$ \\
FOV J1637+1549     & star? (extent$\leq1\farcs4$)     & lobe                      \\
FOV J1655+5015     & aligned galaxy             & lobe                      \\
FOV J1723+5303     & aligned galaxy             & lobe                      \\ 
FOV J2232$-$0903   & not aligned                & lobe                      \\ 
\enddata
\end{deluxetable}

\end{center}

%%%%%%%%%%%%%%%%%%%%%%%%%%%%%%%%%%%%%%%%%%%%%%%%%%%%%%%%%%%%%%
%\clearpage
\begin{center}
\begin{deluxetable}{ll|ll}
\tabletypesize{\normalsize} 
\tablecaption{Two components of the light distribution of spiral
galaxy FOV~J0743+1553
\label{table.optmodel}}
\tablehead{ 
\multicolumn{2}{c}{DeVaucouleurs (bulge)} & \multicolumn{2}{c}{Exponential Disk} }
\startdata
H mag          &\phs   17.44$\pm$0.05     & H mag &\phs   16.20$\pm$0.04      \\
R$_{\rm{eff}}$ &\phs\phn0.27$\pm$0.02$''$ & scale &\phs\phn0.84$\pm$0.02$''$  \\
b/a            &\phs\phn0.71$\pm$0.02     & b/a   &\phs\phn0.25$\pm$0.01      \\
PA             &$-$74.\phn\phn$\pm$2.\degr  & PA    &$-$77.3\phn\phn$\pm$0.1\degr \\

\enddata

\tablecomments{The errors were calculated from the dispersion
of fits using the various sub-exposures and by using 12 different PSF
stars.
}
\end{deluxetable}
\end{center}

\end{document}